\documentclass[sigconf]{acmart}

\usepackage{balance}
\usepackage[english]{babel}
\usepackage{blindtext}
\usepackage{multirow}
\usepackage{xcolor}
\usepackage{colortbl}
\definecolor{alizarin}{rgb}{0.82, 0.1, 0.26}
\definecolor{blizzardblue}{rgb}{0.36, 0.54, 0.66}
\definecolor{blue-violet}{rgb}{0.54, 0.17, 0.89}
\definecolor{darkred}{rgb}{0.54, 0, 0}
\usepackage{adjustbox}
\usepackage{subfigure}

\usepackage{graphicx}
\graphicspath{ {figures/} }

\usepackage{comment}

\usepackage{tabularx}
\usepackage{hyphenat}

\newcommand{\ie}{{\em i.e., }}
\newcommand{\eg}{{\em e.g., }}
\newcommand{\etal}{\textit{et al. }}
\newcommand{\mycomment}[1]{}

\copyrightyear{2024} 
\acmYear{2024} 
\setcopyright{acmlicensed}\acmConference[IMC '24]{Proceedings of the 2024 ACM Internet Measurement Conference}{November 4--6, 2024}{Madrid, Spain}
\acmBooktitle{Proceedings of the 2024 ACM Internet Measurement Conference (IMC '24), November 4--6, 2024, Madrid, Spain}
\acmDOI{10.1145/3646547.3688435}
\acmISBN{979-8-4007-0592-2/24/11}

\setcopyright{none}
\renewcommand\footnotetextcopyrightpermission[1]{}
\begin{document}
\title[Characterizing User Platforms for Video Streaming in Broadband Networks]{Characterizing User Platforms for Video Streaming\\ in Broadband Networks}

\author{Yifan Wang}
\orcid{0009-0009-3472-9821}
\affiliation{%
	\institution{University of New South Wales}
    \city{Sydney}
    \state{NSW}
	\country{Australia}
}
\email{wangyifan.frank@student.unsw.edu.au}

\author{Minzhao Lyu}
\orcid{0000-0001-8677-248X}
\affiliation{%
	\institution{University of New South Wales}
    \city{Sydney}
    \state{NSW}
	\country{Australia}
}
\email{minzhao.lyu@unsw.edu.au}

\author{Vijay Sivaraman}
\orcid{0000-0001-7985-6765}
\affiliation{%
	\institution{University of New South Wales}
    \city{Sydney}
    \state{NSW}
	\country{Australia}
}
\email{vijay@unsw.edu.au}

\ccsdesc[500]{Networks~Network measurement}
\ccsdesc[500]{Networks~Network monitoring}
\ccsdesc[300]{Information systems~Multimedia streaming}
\ccsdesc[300]{Computing methodologies~Machine learning approaches}

\keywords{Network traffic analysis, user platform identification, video streaming, TLS fingerprinting}

\newcommand\todo[1]{\textcolor{red}{#1}}
\newcommand\annotation[1]{\textcolor{blue}{#1}}

\begin{abstract}

Internet Service Providers (ISPs) bear the brunt of being the first port of call for poor video streaming experience. ISPs can benefit from knowing the user's device type (\eg Android, iOS) and software agent (\eg native app, Chrome) to troubleshoot platform-specific issues, plan capacity and create custom bundles. Unfortunately, encryption and NAT have limited ISPs' visibility into user platforms across video streaming providers.
We develop a methodology to identify user platforms for video streams from four popular providers, namely YouTube, Netflix, Disney, and Amazon, by analyzing network traffic in real-time. First, we study the anatomy of the connection establishment process to show how TCP/QUIC and TLS handshakes vary across user platforms. We then develop a classification pipeline that uses 62 attributes extracted from the handshake messages to determine the user device and software agent of video flows with over 96\% accuracy. Our method is evaluated and deployed in a large campus network (mimicking a residential broadband network) serving users including dormitory residents. Analysis of 100+ million video streams over a four-month period reveals insights into the mix of user platforms across the video providers, variations in bandwidth consumption across operating systems and browsers, and differences in peak hours of usage.

\end{abstract}

\maketitle

\section{Introduction} \label{sec:intro}

Broadband Internet Service Providers (ISPs) are often the first to be blamed by users experiencing video freeze or grainy resolution, even when the issue is unrelated to the network. In many instances, the issues arise from the user platform -- choppy video playback on Pixel devices in 2020 was attributed to Android 11 Beta issues \cite{wilde_2020_a}; the YouTube app threw up errors on iOS devices in late 2022 \cite{riyamadaan_2023_youtube}; a software update to Roku devices in 2021 caused intermittent video freeze \cite{perez_2021_roku}; and Hulu deliberately lowered resolution on PC browsers in order to force users to download their proprietary app \cite{villarreal_2020_hulu,dristone_a2023_hulu}.
ISPs, who bear the brunt of customer support calls, can hugely benefit by knowing the user platform, namely device type (iOS or Android smartphone or tablet, Windows or Mac PC, smart TV, Xbox or PlayStation console) and software agent (native app versus a specific browser such as Chrome, Firefox, Safari or Edge) on which a household user is having a poor video streaming experience. This will allow their customer care staff to rapidly filter out known platform-specific issues, prioritize handling of support tickets based on platform prevalence, and issue preemptive advisories to users, all of which can substantially reduce support costs.
 
Visibility into the user platform has other benefits as well for ISPs. The same video watched via a content provider's app may consume significantly higher bandwidth than when watched on a web browser, and these differences can amplify across operating systems \cite{jumani_device-aware_2019,balachandran_developing_2013}. Given that video streaming dominates network traffic, ISP bandwidth provisioning and management models need to account for user device type and software agent heterogeneity, which vary widely from one content provider (\eg YouTube) to another (\eg Netflix). Further, user platform visibility also enables ISPs to perform better customer segmentation (\eg fans who stream live sports via set-top boxes), allowing them to create innovative custom bundles and to run effective up-sell/cross-sell marketing campaigns. Collectively, these initiatives can give an ISP significant competitive advantage, helping reduce costs as well as generate new revenues.
 
Deducing the user platform associated with a streaming session is unfortunately non-trivial for the ISP. Much of the traffic to/from the home today is generally on a single IPv4 address, which is shared among all household devices via network address translation (NAT). While IPv6 may eventually overcome this issue, deployment is immature in many ISPs globally, and is unlikely to displace IPv4 anytime soon. Further, deducing the software agent, such as a browser versus a native app, will require a different technique irrespective of whether the traffic is IPv6 or IPv4, since client-server interactions are predominantly carried within encrypted SSL/TLS sessions today.
 
Much attention has been given to classifying application streams, such as web browsing \cite{panchenko_website_2016,sun_statistical_2002}, video streaming \cite{habibi_gharakheili_itelescope_2019,mazhar_real-time_2018}, video conferencing \cite{michel_enabling_2022}, online gaming \cite{madanapalli_know_2022}, cloud gaming \cite{lyu_network_2024} and metaverse \cite{lyu_metavradar_2023}, but less to identifying the client platform (OS and software agent) associated with each application stream. OS fingerprinting is done in \cite{hagos_machine_2021,shamsi_faulds_2021}, but not coupled with the software agent -- native apps are often optimized differently than browsers for streaming video \cite{spang_Sammy_2023}. Prior works have leveraged distinguishable patterns of TCP-based handshake fields across device firmware \cite{fan_identify_2019,lastovicka_using_2020} and application types \cite{anderson_tls_2019,razaghpanah_studying_2017}, with cipher suites shown to vary from one OS to another \cite{husak_network-based_2015}, and certain TLS handshake fields being unique to certain browsers \cite{anderson_tls_2019}. Unlike prior studies that have a broad focus, our study focuses exclusively on video streams, which constitute over 60\% of Internet traffic. We dive deep into the differences across OSes, browsers, content providers, and their native apps. Further, we consider the presence of QUIC, which is increasingly being used by video content providers -- measurements in tier-1 ISPs indicate that QUIC accounts for nearly 30\% of traffic in EMEA and 16\% in North America \cite{sandvine_quic}. Our methods are therefore more comprehensive in giving ISPs visibility into both device type and software agent for every video stream in their network.
 
Our {\bf first} contribution (\S\ref{sec:analysis}) comprehensively studies the communication process involved in the establishment of a video streaming session across 30 user platforms, \ie combinations of device types and software agents. The device types comprise mobiles, laptops, tablets, PCs and smart TVs running iOS, Android, Windows, macOS, Android TV, etc. For software agents, we consider native apps developed by streaming video providers and browsers such as Chrome, Firefox, Safari and Edge. By collecting and analyzing over 10,000 video flows in our lab from the four major providers, namely YouTube, Netflix, Amazon Prime Video and Disney+, we highlight the variations in their TCP/QUIC and TLS handshake parameters across the 30 user platforms.
 
For our {\bf second} contribution (\S\ref{sec:methodology}), we develop a machine learning-based pipeline to classify the user platform of a video streaming flow. We systematically extract 62 attributes from the initial connection establishment phase of the TCP or QUIC video flow; we then train machine learning models that use different algorithms, hyperparameters and attribute subsets grouped by their computational cost. We demonstrate that our ML models achieve an accuracy of above 96\% in our lab datasets, outperforming prior methods across every video provider considered. 
 
Our {\bf third} contribution (\S\ref{sec:insights}) is the implementation and deployment of our system in a university campus serving tens of thousands of users and a characterization of the video streaming traffic observed across 4 months -- over 100 million video flows spanning 400k hours of watch time.
Our data reveals that mobile devices are more commonly used for YouTube, while PCs/laptops predominate for the other (subscription-based) video services. The latter also seem to demand higher bandwidth, with Amazon Prime Video being the most demanding, especially on Mac PCs that consume 50\% more on average than even smart TVs. 
We believe that our system and the insights collected may be of help for network operators seeking to link video streaming services to capacity planning, user experience troubleshooting and user segmentation. In the spirit of reproducibility and to foster discussion in the research community, we make publicly available the code and training data at \cite{wang_github_2024}.

\section{Related Work} \label{sec:related}
\textbf{OS/application classification:}\label{subsec:client_config}
Classifying operating systems and applications has received considerable attention \cite{callado_survey_2009,finsterbusch_survey_2014,pacheco_towards_2019,papadogiannaki_survey_2021}. 
Prior works that purely use common network and transport layer signatures to classify OS and application have been rendered inaccurate with the growth of complexity in user platforms \cite{lippmann_passive_2003,zalewski_2014_p0f,alan_can_2016,bernaille_early_2006}. Therefore, recent research works leverage certain fields in TLS ClientHello (CHLO) messages to classify device OSes and application types \cite{lastovicka_passive_2022,yang_bayesian_2019}, as such information is directly related to client-side firmware and software configurations for their encryption preferences.
For example, M. Husak \etal \cite{husak_network-based_2015} used cipher suites from TLS CHLO messages to classify web browsers. M. Lastovicka \etal \cite{lastovicka_using_2020} used machine learning models to classify operating systems using 7 features (\eg server name and TLS version) extracted from TLS CHLO messages. B. Anderson \etal \cite{anderson_os_2017} jointly considered current and past TLS CHLO characteristics of a certain device for high-confidence detection of device OS versions. Other works combined TLS CHLO fields with flow statistics (\eg packet size distribution) to classify OS and/or application types \cite{fan_identify_2019,muehlstein_analyzing_2017,shamsimukhametov_is_2022,zheng_mtt_2022,jiang_acdc_2023}.
Compared to prior research, our work focuses on the classification of user platforms (including both device types and software agents), instead of traffic (\ie application type) classification which in our case is based on TLS SNI matching  -- our ML models are activated once the video streaming application is detected. Specifically, we consider per-video streaming flow using TCP/QUIC and TLS handshake fields, without relying on aggregated statistics per device IP which can be rendered ineffective in residential networks with NAT in place. Instead of cherry-picking certain header fields, we evaluate 62 attributes that cover all available handshake fields of a video flow that can vary across user platforms, many of which (particularly for QUIC) are not considered in prior works. Later in \S\ref{sec:benchmark}, we empirically demonstrate the superior performance of our method over prior techniques.

\textbf{Fingerprinting handshake fields:}
A group of prior research works developed tools to measure variations of (TLS) handshake fields that can exist in different device types (\eg client or server), applications (\eg browsers or social media apps), and vulnerable firmware and software configurations. For such purposes, JA3 \cite{althouse_2023_ja3} has been developed as a popular tool for fingerprinting TLS CHLO fields from a network device, which has been extended to include server-side fingerprinting in JA3S \cite{althouse_2019_ja3s}.
A. Razaghpanah \etal \cite{razaghpanah_studying_2017} developed a tool for extracting certain TLS CHLO fields from network traffic from Android devices, which are useful in identifying those with security vulnerabilities or misconfigurations.
B. Anderson \etal \cite{anderson_tls_2019} proposed a system that extracts TLS CHLO messages from standard operational environment devices in an enterprise network as an effective monitoring measure for the running processes on each host and their TLS configurations.
M. Sosnowski \etal \cite{brunstrom_dissectls_2023} evaluated five popular server-side TLS fingerprinting methods and developed an active server TLS scanner.
Prior works only focused on certain handshake fields in TCP-based TLS flows.
In this paper, we develop a system that automatically extracts and formalizes all available handshake fields from not only TCP-based TLS but also QUIC-based TLS for video flows.

\textbf{Traffic analysis of streaming video services:}\label{subsec:video_traffic}
There is a large body of prior research works that analyze network traffic for streaming video services to provide visibility into the usage and user experience of video sessions for network operators. 
Machine learning algorithms that consume statistical attributes extracted from video flows such as packet inter-arrival times, downstream/upstream packet rates and throughput are extensively used by these works. 
Their prediction objectives include detecting video flows \cite{habibi_gharakheili_itelescope_2019,dong_novel_2017}, classifying content providers \cite{madanapalli_reclive_2021,dias_innovative_2019}, modeling video adaptation behaviors \cite{mondal_candid_2017,xu_csi_2020}, and inferring the status of streaming experience such as resolution, stall, and startup delay \cite{dimopoulos_measuring_2016,gutterman_requet_2019,wassermann_vicrypt_2020,afzal_characterization_2017}.
An important aspect that has not been well captured by prior works is identifying user platforms of video flows, which can influence the observed streaming behavior and the user experience \cite{li_quantifying_2018}.
In this paper, we develop a lightweight method focused on responsively characterizing user platforms for each streaming video flow using only handshake messages prior to the delivery of actual video content.

\begin{figure}[t!]
	\centering
	\includegraphics[width=\linewidth]{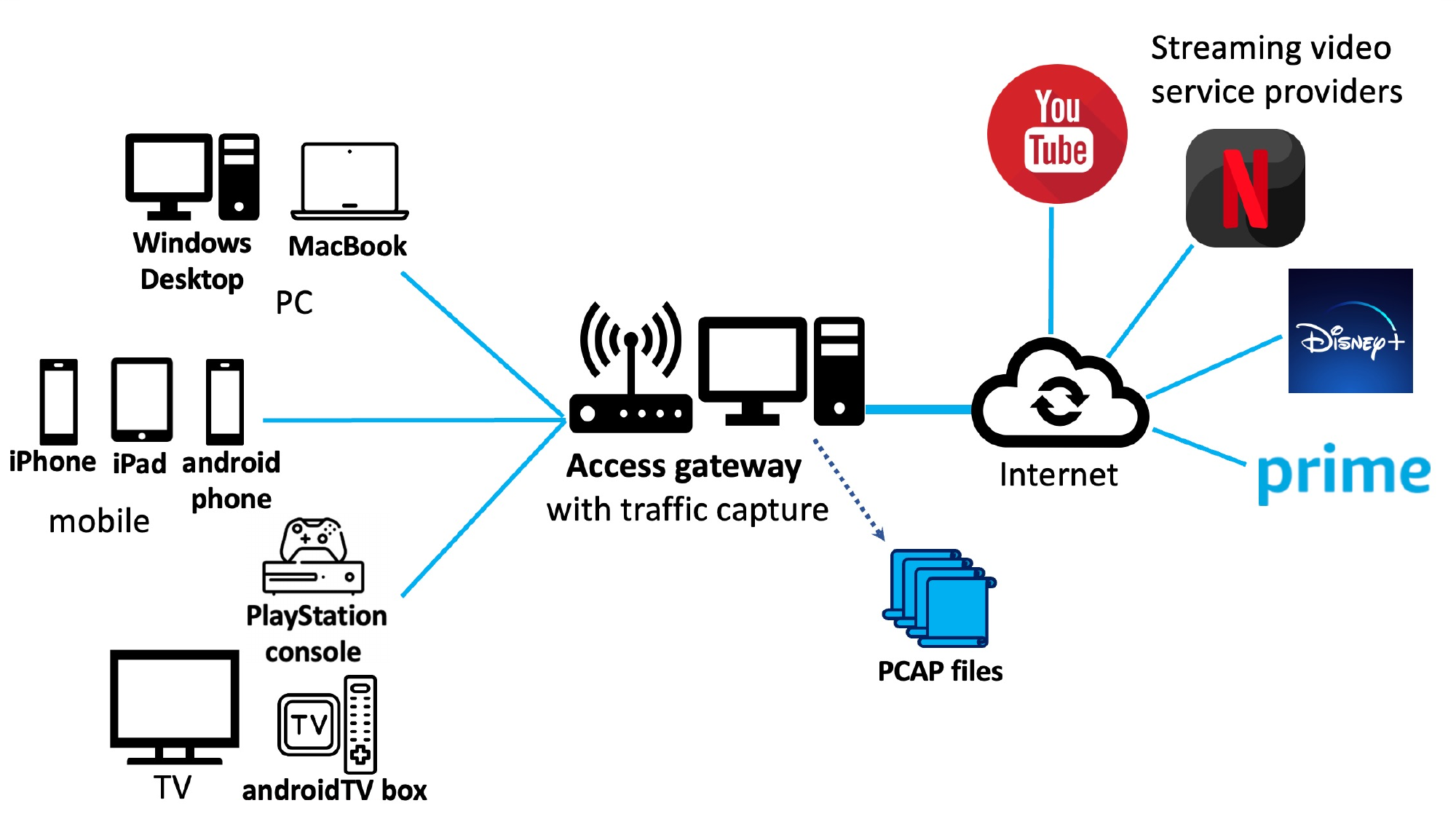}
	\caption{Experimental setup for video streaming traffic trace collection.}
	\label{fig:lab_setup}
\end{figure}

\section{Handshake Characteristics of Video Streaming Sessions} \label{sec:analysis}
In this section, we describe our experimental setup for capturing traffic traces (\S\ref{sec:analysis-dataset}), then systematically study the communication process involved in establishing video streaming sessions using a variety of device types and software agents (\S\ref{sec:analysis-anatomy}), and finally identify the handshake fields that take on different values across user platforms and video content providers (\S\ref{sec:analysis-handshake}).

\subsection{Experimental Setup and Dataset}\label{sec:analysis-dataset}
In Fig.~\ref{fig:lab_setup}, we show our experimental framework to collect traffic trace files (\ie PCAPs) for video streaming sessions from the four major video content providers, namely YouTube, Netflix, Amazon Prime Video and Disney+. The setup consists of 3 mobile devices (iPhone, iPad, Android phone) running iOS and Android operating systems, 2 Windows desktop PCs, 2 macOS MacBooks, 2 smart TVs (including one with a built-in Android TV system and the other one connected to an Android TV set-top box), and a PlayStation gaming console. Having a variety of different devices expedites data collection with multiple users streaming videos concurrently. These devices connect via Wi-Fi to the access gateway for Internet access, from which we collect PCAP files using the Wireshark tool.

We capture traffic traces containing more than 100 video sessions for each of the 30 user platforms in our lab, comprising various device types and software agents. Our dataset contains 17 unique types of user platforms as we have redundancy for some OS types such as 2 Windows and 2 macOS. Each session is composed of one or multiple video flows, for a total of nearly 10,000 flows. We show the detailed composition of our dataset in Table~\ref{tab:dataset}. Some of the devices run the same OS, thus are considered as being the same from a user platform perspective. For example, both iPhone and iPad in our setup are using iOS, and so are captured under a single (Mobile, iOS) category in Table~\ref{tab:dataset}. The duration of each session was at least one minute, sufficient to capture all the handshake messages exchanged between the user device and the server of the respective provider. Note that we are not interested in capturing packets containing any video content payload. The devices were running the most recent versions of operating systems, browsers and apps. Some browsers, \eg Chrome, Firefox and Edge on Windows/Mac PCs, allow users to configure the transport layer protocol as either QUIC or TCP, which can impact the connection establishment messages exchanged. Our dataset encompasses all these different scenarios and has comprehensive coverage across all different configuration options. Our dataset is available on our university cloud storage platform and can be shared upon request, as detailed in \cite{wang_github_2024}.
	
	\begin{table}[t!]
	\centering
	\caption{Number of video flows per content provider, \ie \textbf{\color{alizarin}YouTube (YT)}, \textbf{\color{blue-violet}Netflix (NF)}, \textbf{\color{blue}Amazon Prime Video (AP)} or \textbf{\color{blizzardblue}Disney+ (DN)} captured for each combination of device type and software agent in our collected traffic traces. A user platform not supported by the content provider is marked as ---.}
	\fontsize{7.8pt}{10.5pt}\selectfont
	\begin{tabular}{|l|l|l|cccc|}
		\hline
		\rowcolor[HTML]{EFEFEF} 
		\cellcolor[HTML]{EFEFEF}                                  & \cellcolor[HTML]{EFEFEF}                                   & \cellcolor[HTML]{EFEFEF}                                          & \multicolumn{4}{c|}{\cellcolor[HTML]{EFEFEF}\textbf{Number of video flows}}                                                                                                                  \\ \cline{4-7} 
		\rowcolor[HTML]{EFEFEF} 
		\multirow{-2}{*}{\cellcolor[HTML]{EFEFEF}\textbf{Device}} & \multirow{-2}{*}{\cellcolor[HTML]{EFEFEF}\textbf{OS Type}} & \multirow{-2}{*}{\cellcolor[HTML]{EFEFEF}\textbf{Software Agent}} & \multicolumn{1}{c|}{\cellcolor[HTML]{EFEFEF}\textbf{\color{alizarin}YT}} & \multicolumn{1}{c|}{\cellcolor[HTML]{EFEFEF}\textbf{\color{blue-violet}NF}} & \multicolumn{1}{c|}{\cellcolor[HTML]{EFEFEF}\textbf{\color{blizzardblue}DN}} & \textbf{\color{blue}AP} \\ \hline
		&                                                            & Chrome                                                            & \multicolumn{1}{c|}{\color{alizarin}411}                                 & \multicolumn{1}{c|}{\color{blue-violet}202}                                 & \multicolumn{1}{c|}{\color{blizzardblue}199}                                 & \color{blue}215         \\ \cline{3-7} 
		&                                                            & Edge                                                              & \multicolumn{1}{c|}{\color{alizarin}406}                                 & \multicolumn{1}{c|}{\color{blue-violet}208}                                 & \multicolumn{1}{c|}{\color{blizzardblue}200}                                 & \color{blue}200         \\ \cline{3-7} 
		&                                                            & Firefox                                                           & \multicolumn{1}{c|}{\color{alizarin}466}                                 & \multicolumn{1}{c|}{\color{blue-violet}207}                                 & \multicolumn{1}{c|}{\color{blizzardblue}204}                                 & \color{blue}195         \\ \cline{3-7} 
		& \multirow{-4}{*}{Windows}                                  & Native app                                                        & \multicolumn{1}{c|}{\color{alizarin}---}                                 & \multicolumn{1}{c|}{\color{blue-violet}204}                                 & \multicolumn{1}{c|}{\color{blizzardblue}211}                                 & \color{blue}186         \\ \cline{2-7} 
		&                                                            & Safari                                                            & \multicolumn{1}{c|}{\color{alizarin}200}                                 & \multicolumn{1}{c|}{\color{blue-violet}204}                                 & \multicolumn{1}{c|}{\color{blizzardblue}200}                                 & \color{blue}201         \\ \cline{3-7} 
		&                                                            & Chrome                                                            & \multicolumn{1}{c|}{\color{alizarin}407}                                 & \multicolumn{1}{c|}{\color{blue-violet}213}                                 & \multicolumn{1}{c|}{\color{blizzardblue}202}                                 & \color{blue}208         \\ \cline{3-7} 
		&                                                            & Edge                                                              & \multicolumn{1}{c|}{\color{alizarin}402}                                 & \multicolumn{1}{c|}{\color{blue-violet}204}                                 & \multicolumn{1}{c|}{\color{blizzardblue}202}                                 & \color{blue}210         \\ \cline{3-7} 
		\multirow{-7}{*}{PC}                                      &                                                            & Firefox                                                           & \multicolumn{1}{c|}{\color{alizarin}467}                                 & \multicolumn{1}{c|}{\color{blue-violet}212}                                 & \multicolumn{1}{c|}{\color{blizzardblue}202}                                 & \color{blue}199         \\ \cline{3-7} 
		& \multirow{-5}{*}{macOS}                                    & Native app                                                        & \multicolumn{1}{c|}{\color{alizarin}---}                                 & \multicolumn{1}{c|}{\color{blue-violet}---}                                 & \multicolumn{1}{c|}{\color{blizzardblue}---}                                 & \color{blue}200         \\ \hline
		&                                                            & Chrome                                                            & \multicolumn{1}{c|}{\color{alizarin}107}                                 & \multicolumn{1}{c|}{\color{blue-violet}---}                                 & \multicolumn{1}{c|}{\color{blizzardblue}---}                                 & \color{blue}---         \\ \cline{3-7} 
		&                                                            & Samsung Internet                                                  & \multicolumn{1}{c|}{\color{alizarin}103}                                 & \multicolumn{1}{c|}{\color{blue-violet}---}                                 & \multicolumn{1}{c|}{\color{blizzardblue}---}                                 & \color{blue}---         \\ \cline{3-7} 
		& \multirow{-3}{*}{Android}                                  & Native app                                                        & \multicolumn{1}{c|}{\color{alizarin}100}                                 & \multicolumn{1}{c|}{\color{blue-violet}102}                                 & \multicolumn{1}{c|}{\color{blizzardblue}106}                                 & \color{blue}111         \\ \cline{2-7} 
		&                                                            & Safari                                                            & \multicolumn{1}{c|}{\color{alizarin}203}                                 & \multicolumn{1}{c|}{\color{blue-violet}---}                                 & \multicolumn{1}{c|}{\color{blizzardblue}---}                                 & \color{blue}---         \\ \cline{3-7} 
		&                                                            & Chrome                                                            & \multicolumn{1}{c|}{\color{alizarin}213}                                 & \multicolumn{1}{c|}{\color{blue-violet}---}                                 & \multicolumn{1}{c|}{\color{blizzardblue}---}                                 & \color{blue}---         \\ \cline{3-7} 
		\multirow{-6}{*}{Mobile}                                  & \multirow{-3}{*}{iOS}                                      & Native app                                                        & \multicolumn{1}{c|}{\color{alizarin}203}                                 & \multicolumn{1}{c|}{\color{blue-violet}215}                                 & \multicolumn{1}{c|}{\color{blizzardblue}306}                                 & \color{blue}372         \\ \hline
		& Android TV                                                 & Native app                                                        & \multicolumn{1}{c|}{\color{alizarin}200}                                 & \multicolumn{1}{c|}{\color{blue-violet}116}                                 & \multicolumn{1}{c|}{\color{blizzardblue}107}                                 & \color{blue}113         \\ \cline{2-7} 
		\multirow{-2}{*}{TV}                                      & PlayStation                                                & Native app                                                        & \multicolumn{1}{c|}{\color{alizarin}105}                                 & \multicolumn{1}{c|}{\color{blue-violet}100}                                 & \multicolumn{1}{c|}{\color{blizzardblue}100}                                 & \color{blue}103         \\ \hline
	\end{tabular}
	\label{tab:dataset}
\end{table}

\subsection{Anatomy of Video Streaming Sessions}\label{sec:analysis-anatomy}
We now discuss the typical communication process involved in establishing a video streaming session between a client device and a streaming server.

\subsubsection{Communication process}
A streaming video session essentially consists of two sequential stages, \ie initialization and playback. The initialization stage, comprising connection setup handshake exchanges, alongside sharing of relevant metadata information, has received little attention from the community; later we will show its importance in characterizing user platforms. The playback stage on the other hand, comprising streaming of the video itself, has been extensively studied for detecting video traffic streams, quantifying QoE, and so on, as summarized in \S\ref{sec:related}. 

A flow diagram depicting the two stages is shown in Fig.~\ref{fig:youtube_concept_overview}.
In the initialization stage, the client device interacts with a management server specifying the service request, user configurations and connection parameters such as the device type, software agent used and supported network protocols; step \textcircled{1} in Fig.~\ref{fig:youtube_concept_overview}. Then, in step \textcircled{2}, the management server responds with the streaming information (\eg URLs of the content servers and video/audio formats) along with control parameters (\eg media player configurations) to be adapted by the software agent. In step \textcircled{3}, the actual video playback process starts after the client requests video and audio data from the content server located using the URL information acquired in the previous step. This request also specifies video quality metrics such as resolution which can be adjusted at any time during the playback process, either manually or dynamically by the client-side player depending on the network conditions and playback quality. In step \textcircled{4}, the video and audio data are streamed to the client. The software agent on it may periodically send playback status information to a management server, as indicated in \textcircled{5}, to help the content provider keep track of service usage, session status, video quality and the like. 

Steps \textcircled{1} to \textcircled{4} exist in all video streaming traces we have collected across the different providers, whereas step \textcircled{5} is only observed in certain video sessions such as on macOS devices watching YouTube on a Chrome browser.

\subsubsection{Anatomy of network communication}
\begin{figure}[!t]
	\mbox{
		\hspace{-4mm}
		\subfigure[Communication process.]{
			{\includegraphics[width=0.4\linewidth]{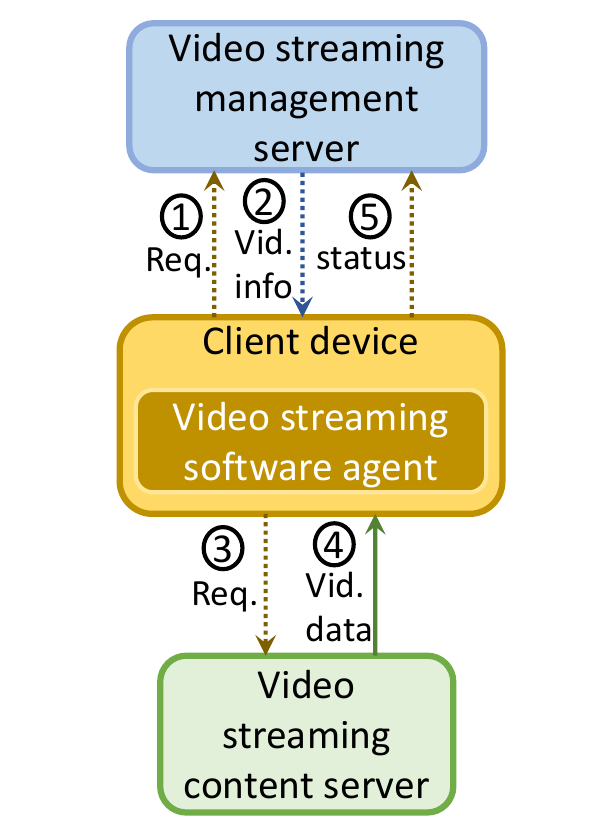}}\quad
			\label{fig:youtube_concept_overview}
		}
		\hspace{-10mm}
		\subfigure[Practical example using YouTube.]{
			{\includegraphics[width=0.7\linewidth]{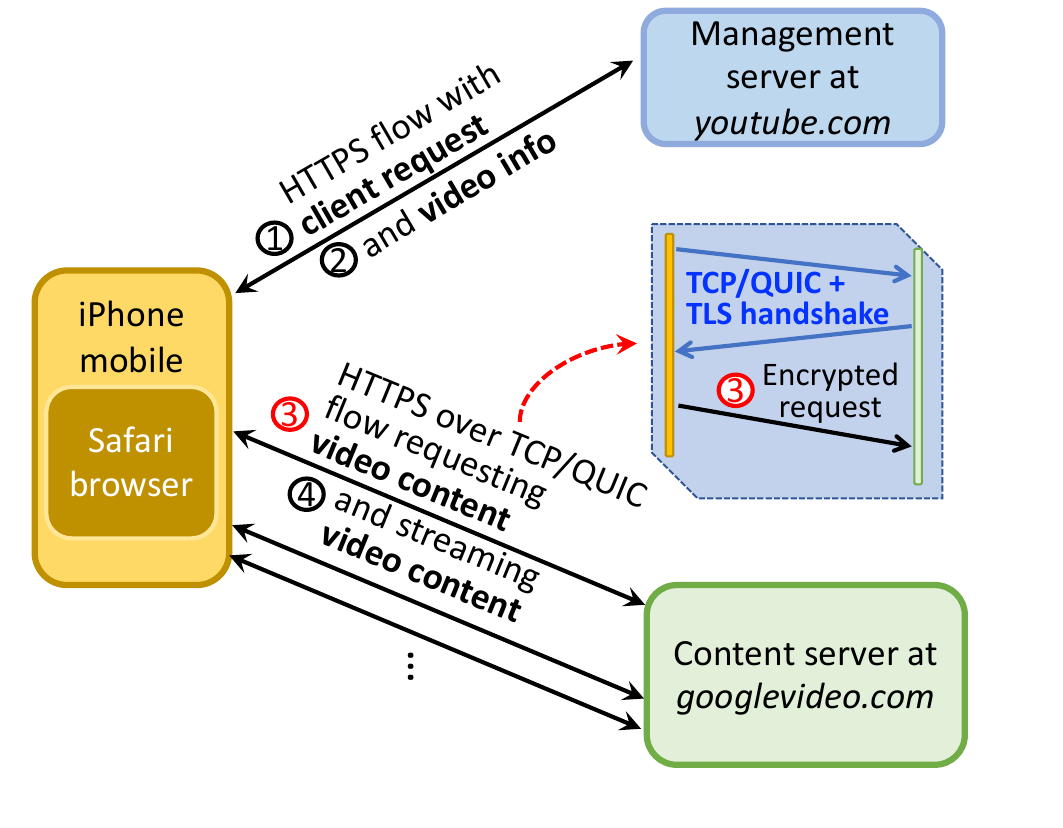}}\quad
			\label{fig:youtube_anatomy}
		}								
	}
	\caption{Anatomy of network communication for a video streaming session.}
	\label{fig:processAndAnatomy}
\end{figure}

From the perspective of network communication, the above steps are carried by HTTPS flows over either TCP or QUIC as the transport layer protocol. To illustrate the detailed communication process, Fig.~\ref{fig:youtube_anatomy} shows an example of a YouTube session operated from an iPhone using Safari browser.
Video sessions on other device types, software agents and content providers share a similar anatomy, omitted for brevity. The steps circled in Fig.~\ref{fig:youtube_anatomy} correspond to those shown in Fig.~\ref{fig:youtube_concept_overview}.

The initial client request flow, \ie step \textcircled{1}, is always sent via a single HTTPS flow to a management server, typically \textit{youtube.com}, \textit{netflix.com}, \textit{primevideo.com}, \textit{disneyplus.com} for the four content providers, respectively. For YouTube, the flow is either carried by TCP or QUIC depending on the client configuration, whereas the other three providers, \ie Netflix, Amazon and Disney+ use only TCP.
In step \textcircled{2},  the management server sends streaming information and player configuration to the client on the same HTTPS flow. However, if the client is configured to pre-load video metadata on a menu page, rather than requesting it dynamically from the server, then this information could be carried over multiple subsequent HTTPS flows. 

During the video playback stages, steps \textcircled{3} and \textcircled{4}, the client fetches the video from a content server (\eg \textit{googlevideo.com} for YouTube) using one or more HTTPS flows over either TCP or QUIC. We have observed three scenarios in this process. One, in which playback is delivered by a single HTTPS flow containing both video and audio data. Two, in which playback is delivered over multiple concurrent HTTPS flows carrying video and audio. Three, in which multiple HTTPS flows are activated in different time slots, each delivering chunks of video and audio data. For example, in some YouTube sessions we have collected, there are flows that send several chunks of video and audio data in the first few seconds (\eg 3 seconds) and then go idle while the remaining video and audio data is streamed by another flow. Since client information and playback data are all encrypted by TLS, network operators only have visibility into the TCP/QUIC and TLS handshake messages, as visually shown in the blue region in Fig.~\ref{fig:youtube_anatomy}, and volumetric information of the payloads.

Next, we show that handshake messages carried by the first few (< 5) connection establishment packets contain fields that are strong indicators to characterize user platforms.

\subsection{Handshake Characteristics Across User Platforms}\label{sec:analysis-handshake}
The flows that stream video content are first initialized by a series of handshake messages. By analyzing the trace files, we find that information within these handshake messages is highly correlated with the OS, software agent and the provider of video content. In the following, we systematically discuss these properties. 

\begin{figure*}[t!]
	\includegraphics[width=\linewidth]{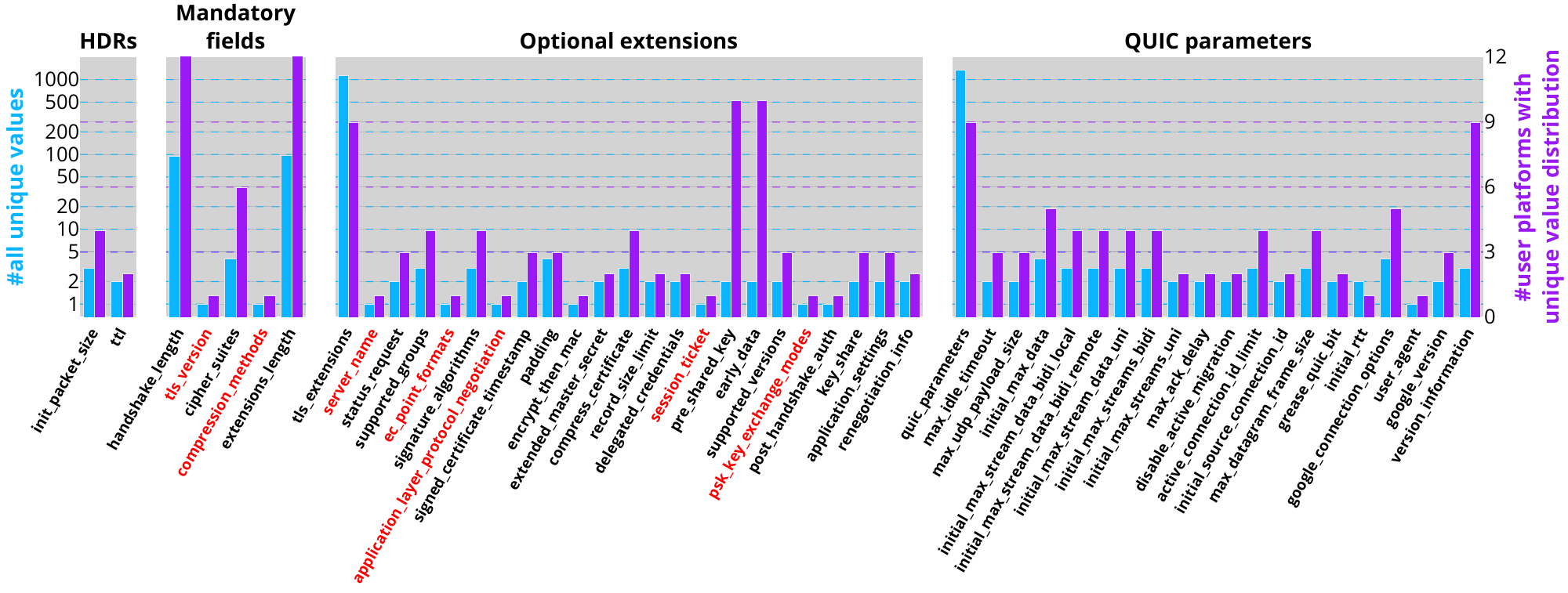}
	\vspace{-4mm}
	\caption{Number of unique values (left blue in log scale) and number of user platforms with different value distributions (right purple in linear scale) for each handshake field in YouTube flows over QUIC.}
	\label{fig:field_distribution_quic}
\end{figure*}

\subsubsection{Handshake fields and their categorization}
A video flow delivered by HTTPS has two handshake processes: one for the transport layer protocol (\ie TCP or QUIC) and the other for TLS encryption, both of which occur prior to the encrypted video content being streamed. 

\paragraph{Transport layer handshake}
Both TCP and QUIC require a handshake process to establish connections. A TCP three-way handshake contains TCP header flags and options, such as window size, selective acknowledgment and max segment size, which are set by the user device and the software agent. We ignore most of the TCP flags in handshake messages such as SYN, ACK, RST and FIN as they do not differ across user platforms. Exceptions are TCP CWR and ECE, which are related to congestion control policies used by the device OS or software agent \cite{rfc3168}.

QUIC is designed to reduce the connection setup latency overhead and so its handshake via the first flow packet (\ie QUIC Initial packet) is integrated with the TLS handshake, as discussed below. In addition, we find that the time-to-live field and packet size in the IP header of the Initial packet are often correlated with the device type, as reported by \cite{sk_identify_2021}.

\paragraph{TLS handshake}
TLS handshake consists of a ClientHello followed by a ServerHello and subsequent encryption negotiations, and is executed right after the TCP three-way handshake or along with the first QUIC flow packet. The ClientHello contains customized information provided by the user device and is highly correlated with the device type and software agent that plays the video. We decompose the TLS handshake fields into three categories. 

The first category is called \textbf{mandatory fields}. These fields always appear in the ClientHello of streaming video flows regardless of the underlying transport layer protocol (TCP or QUIC) and the specifications of user devices, OSes and software agents. These include handshake length, TLS version, cipher suites and compression methods. 

The second category is \textbf{optional extensions}. These fields only appear in video flows as defined by the logic embedded in a device OS and software agent. In our dataset, a given user platform typically uses unique combinations of optional extensions, each set to a specific value. For example, Firefox browsers running on Windows and macOS PCs typically set the value of \textit{record\_size\_limit} extension to 16385, whereas other user platforms do not use this extension.

In addition to the above categories, there are \textbf{parameters} in the ClientHello that are specifically available for video flows over \textbf{QUIC}. These parameters are contained in the collection \textit{quic\_transport\_parameters} with extension code 57 \cite{rfc9001,internetassignednumbersauthority_2023_transport}, which are set for specific QUIC preferences in connection establishment of certain user platforms. For example, Firefox browsers on Windows desktop PCs use the parameter \textit{grease\_quic\_bit} to indicate its deprecation of certain flag bits in QUIC headers \cite{rfc9287}.

\subsubsection{Handshake fields across user platforms}	\label{subsec:handshake_analysis}
We now analyze the distribution of values contained in the fields of various TCP/QUIC and TLS handshake messages. 
The aim is to demonstrate the similarities and differences in the values contained within these fields across user platforms, which will form the basis of our machine learning model. We note that some fields are not numerical but categorical or lists, such as the mandatory fields \textit{tls\_version} and \textit{cipher\_suites} in TLS CHLO and \textit{supported\_groups}, \textit{signature\_algorithms} in TLS optional extensions. To simplify the analysis, we convert the values contained in such fields to integers by a 1:1 mapping between the values contained in the fields to a unique number. For instance, in our dataset, the field \textit{compress\_certificate} takes 2 values when carried over QUIC, \ie \emph{zlib} and \emph{brotli}, which are uniquely mapped as 1 and 2 for the purposes of our analysis. Therefore, a video flow containing \emph{zlib} as the value for \textit{compress\_certificate} is represented as 1 in our dataset. If a field does not appear in a flow, a value of 0 is assigned to it. 

For each handshake field in YouTube QUIC video flows, we show the number of unique values as a blue bar in Fig.~\ref{fig:field_distribution_quic} (in log scale), while the number of user platforms having a unique value distribution compared to their counterparts for this particular field is shown as a purple bar (in linear scale). 
There are 7 fields that only have one unique value regardless of user platforms. These are highlighted as red labels in Fig.~\ref{fig:field_distribution_quic}. Apparently, those fields are useless in differentiating user platforms for YouTube QUIC video flows.
However, four of these fields, \ie \textit{ec\_point\_formats}, \textit{ALPN}, \textit{session\_ticket} and \textit{psk\_key\_exchange\_modes} take different values across user platforms for TCP flows, thus, can serve as useful indicators for TCP scenarios.
We provide two heatmaps as Appendix~\ref{sec:Appendix-Handshake} Fig.~\ref{fig:heatmap_quic} and Fig.~\ref{fig:heatmap_tcp} to show the median values of all handshake fields in both TCP and QUIC YouTube video flows respectively, rather than just their counts.

As detailed in Appendix~\ref{sec:Appendix-Handshake}, similar conclusions on value distributions of handshake fields can be reached for other three video providers we have studied in this work, which are not explicitly discussed here for simplicity.
In the next section, aiming for better classification performance, we systematically evaluate the importance of formalized attributes extracted from those handshake fields.

\begin{figure*}[t!]
	\includegraphics[width=0.9\linewidth]{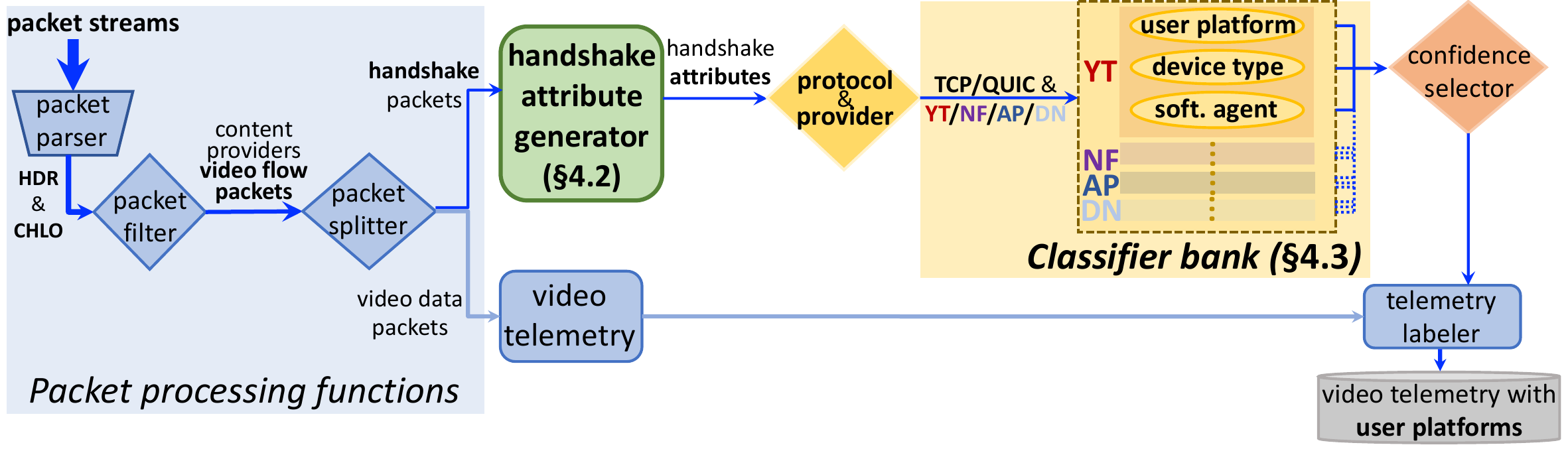}
	\vspace{-2mm}
	\caption{Packet processing pipeline for classification of video streaming user platforms.}
	\label{fig:pipeline}
	\vspace{-3mm}
\end{figure*}

\section{Classifying User Platforms for Video Streams in Real-Time} \label{sec:methodology}
In this section, we first develop a classification pipeline (\S\ref{sec:classification-method}), then identify the fields that are relevant for classification purposes (\S\ref{sec:classification-attributes}), and finally evaluate the efficacy of three machine learning algorithms to achieve our desired objectives (\S\ref{sec:classification-models}).

\subsection{Packet Processing Pipeline}\label{sec:classification-method}
We develop a generalized packet processing pipeline to classify user platforms for each streaming video flow and apply it to the four providers considered in this paper. The schematic of the pipeline is shown in Fig.~\ref{fig:pipeline}.

The pipeline first takes raw packet streams as input, and parses them to identify video flows that belong to the four providers using port numbers and service names extracted from unencrypted packet headers (HDR) and ClientHello (CHLO) SNIs. These packets are further split into handshake packets for classification and payload packets for telemetry, \eg to obtain session duration, volume and throughput. This forms the preprocessing stage. 

Next, each handshake packet is processed to extract attributes that can be readily fed into three machine learning classifiers that predict the composite user platform, device type and software agent for the respective streaming service provider. Therefore, in our later implementation that analyzes four video streaming providers, we have twelve classifiers (\ie three classifiers per provider) in total.
The code for extracting handshake attributes from Client Hello packets is made publicly available at \cite{wang_github_2024}. 
In the small minority of cases where the confidence (\ie probability of the predicted class) in predicting the composite user platform is $<80\%$, we consider predicting the device type (\ie OS) and software agent (\ie native app or specific browser) individually, to have high confidence in classifying either of them accurately, so that at least partial information related to the user platform can be predicted confidently.
The predicted user platform for each video flow can then be correlated using flow metadata and timestamps with real-time telemetry for further insight analysis.
If none of the classifiers can produce an inference result with $>80\%$ confidence, we reject the classification and determine the video flow to be from an unknown user platform. Further classifying the unknown user platform requires respective ground-truth training data to be collected, thus, is not within the scope of this paper.

\subsection{Attributes Derived from Handshake Fields} \label{sec:classification-attributes}
We identify 62 fields of interest, of which 20 are numerical, 31 categorical, and 11 of type list. The detailed specifications of these attributes, applicable to both QUIC and TCP video flows, are provided in Table~\ref{tab:AppendixFeatureList}.

 \begin{table*}[h!]
	\centering
	\caption{Handshake fields of video flows and formalized attributes.}
	\vspace{-3mm}
	\resizebox{1.65\columnwidth}{!}{\begin{tabular}{|l|l|l|l|l|l|}
			\hline
			\rowcolor[HTML]{C0C0C0} 
			\textbf{Handshake field}                & \textbf{Transport protocol} & \textbf{Category}   & \textbf{Attribute label} & \textbf{Attribute type} & \textbf{Attribute cost} \\ \hline
			init\_packet\_size                         & TCP, QUIC                  & transport layer     & $t_1$                       & numerical               & low                     \\ \hline
			ttl                                       & TCP, QUIC                   & transport layer     & $t_2$                       & numerical               & low                     \\ \hline
			tcp\_cwr                                  & TCP                         & transport layer     & $t_3$                       & presence                & low                     \\ \hline
			tcp\_ece                                  & TCP                         & transport layer     & $t_4$                       & presence                & low                     \\ \hline
			tcp\_urg                                  & TCP                         & transport layer     & $t_5$                       & presence                & low                     \\ \hline
			tcp\_ack                                  & TCP                         & transport layer     & $t_6$                       & presence                & low                     \\ \hline
			tcp\_psh                                  & TCP                         & transport layer     & $t_7$                       & presence                & low                     \\ \hline
			tcp\_rst                                  & TCP                         & transport layer     & $t_8$                       & presence                & low                     \\ \hline
			tcp\_syn                                  & TCP                         & transport layer     & $t_9$                       & presence                & low                     \\ \hline
			tcp\_fin                                  & TCP                         & transport layer     & $t_{10}$                      & presence                & low                     \\ \hline
			tcp\_window\_size                         & TCP                         & transport layer     & $t_{11}$                      & numerical               & low                     \\ \hline
			tcp\_mss                                  & TCP                         & transport layer     & $t_{12}$                      & numerical               & low                     \\ \hline
			tcp\_window\_scale                        & TCP                         & transport layer     & $t_{13}$                      & numerical               & low                     \\ \hline
			tcp\_sack\_permitted                      & TCP                         & transport layer     & $t_{14}$                      & presence                & low                     \\ \hline
			\rowcolor[HTML]{ECF4FF} 
			handshake\_length                         & TCP, QUIC                   & mandatory fields    & $m_1$                       & numerical               & low                     \\ \hline
			\rowcolor[HTML]{ECF4FF} 
			tls\_version                              & TCP, QUIC                   & mandatory fields    & $m_2$                       & categorical             & medium                  \\ \hline
			\rowcolor[HTML]{ECF4FF} 
			cipher\_suites                            & TCP, QUIC                   & mandatory fields    & $m_3$                       & list                    & high                    \\ \hline
			\rowcolor[HTML]{ECF4FF} 
			compression\_methods                      & TCP, QUIC                   & mandatory fields    & $m_4$                       & length                  & low                     \\ \hline
			\rowcolor[HTML]{ECF4FF} 
			extensions\_length                      & TCP, QUIC                   & mandatory fields    & $m_5$                       & numerical                  & low                     \\ \hline
			\rowcolor[HTML]{DAE8FC} 
			tls\_extensions                           & TCP, QUIC                   & optional extensions & $o_1$                       & list                    & high                    \\ \hline
			\rowcolor[HTML]{DAE8FC} 
			server\_name                              & TCP, QUIC                   & optional extensions & $o_2$                       & length                  & low                     \\ \hline
			\rowcolor[HTML]{DAE8FC} 
			status\_request                           & TCP, QUIC                   & optional extensions & $o_3$                       & categorical             & medium                  \\ \hline
			\rowcolor[HTML]{DAE8FC} 
			supported\_groups                         & TCP, QUIC                   & optional extensions & $o_4$                       & list                    & high                    \\ \hline
			\rowcolor[HTML]{DAE8FC} 
			ec\_point\_formats                        & TCP, QUIC                   & optional extensions & $o_5$                       & categorical             & medium                  \\ \hline
			\rowcolor[HTML]{DAE8FC} 
			signature\_algorithms                     & TCP, QUIC                   & optional extensions & $o_6$                       & list                    & high                    \\ \hline
			\rowcolor[HTML]{DAE8FC} 
			application\_layer\_protocol\_negotiation & TCP, QUIC                   & optional extensions & $o_7$                       & list                    & high                    \\ \hline
			\rowcolor[HTML]{DAE8FC} 
			signed\_certificate\_timestamp            & TCP, QUIC                   & optional extensions & $o_8$                       & length                  & low                     \\ \hline
			\rowcolor[HTML]{DAE8FC} 
			padding                                   & TCP, QUIC                   & optional extensions & $o_9$                       & length                  & low                     \\ \hline
			\rowcolor[HTML]{DAE8FC} 
			encrypt\_then\_mac                        & TCP, QUIC                   & optional extensions & $o_{10}$                      & presence                & low                     \\ \hline
			\rowcolor[HTML]{DAE8FC} 
			extended\_master\_secret                  & TCP, QUIC                   & optional extensions & $o_{11}$                      & presence                & low                     \\ \hline
			\rowcolor[HTML]{DAE8FC} 
			compress\_certificate                     & TCP, QUIC                   & optional extensions & $o_{12}$                      & categorical             & medium                  \\ \hline
			\rowcolor[HTML]{DAE8FC} 
			record\_size\_limit                       & TCP, QUIC                   & optional extensions & $o_{13}$                      & numerical               & low                     \\ \hline
			\rowcolor[HTML]{DAE8FC} 
			delegated\_credentials                    & TCP, QUIC                   & optional extensions & $o_{14}$                      & list                    & high                    \\ \hline
			\rowcolor[HTML]{DAE8FC} 
			session\_ticket                           & TCP, QUIC                   & optional extensions & $o_{15}$                      & length                  & low                     \\ \hline
			\rowcolor[HTML]{DAE8FC} 
			pre\_shared\_key                          & TCP, QUIC                   & optional extensions & $o_{16}$                      & presence                & low                     \\ \hline
			\rowcolor[HTML]{DAE8FC} 
			early\_data                               & TCP, QUIC                   & optional extensions & $o_{17}$                      & length                  & low                     \\ \hline
			\rowcolor[HTML]{DAE8FC} 
			supported\_versions                       & TCP, QUIC                   & optional extensions & $o_{18}$                      & list                    & high                    \\ \hline
			\rowcolor[HTML]{DAE8FC} 
			psk\_key\_exchange\_modes                 & TCP, QUIC                   & optional extensions & $o_{19}$                      & categorical             & medium                  \\ \hline
			\rowcolor[HTML]{DAE8FC} 
			post\_handshake\_auth                     & TCP, QUIC                   & optional extensions & $o_{20}$                      & presence                & low                     \\ \hline
			\rowcolor[HTML]{DAE8FC} 
			key\_share                                & TCP, QUIC                   & optional extensions & $o_{21}$                      & list                    & high                    \\ \hline
			\rowcolor[HTML]{DAE8FC} 
			application\_settings                     & TCP, QUIC                   & optional extensions & $o_{22}$                      & list                    & high                    \\ \hline
			\rowcolor[HTML]{DAE8FC} 
			renegotiation\_info                       & TCP, QUIC                   & optional extensions & $o_{23}$                      & presence                & low                     \\ \hline
			\rowcolor[HTML]{C6D8F0} 
			quic\_parameters                          & QUIC                        & QUIC parameters     & $q_1$                       & list                    & high                    \\ \hline
			\rowcolor[HTML]{C6D8F0} 
			max\_idle\_timeout                        & QUIC                        & QUIC parameters     & $q_2$                       & numerical               & low                     \\ \hline
			\rowcolor[HTML]{C6D8F0} 
			max\_udp\_payload\_size                   & QUIC                        & QUIC parameters     & $q_3$                       & numerical               & low                     \\ \hline
			\rowcolor[HTML]{C6D8F0} 
			initial\_max\_data                        & QUIC                        & QUIC parameters     & $q_4$                       & numerical               & low                     \\ \hline
			\rowcolor[HTML]{C6D8F0} 
			initial\_max\_stream\_data\_bidi\_local   & QUIC                        & QUIC parameters     & $q_5$                       & numerical               & low                     \\ \hline
			\rowcolor[HTML]{C6D8F0} 
			initial\_max\_stream\_data\_bidi\_remote  & QUIC                        & QUIC parameters     & $q_6$                       & numerical               & low                     \\ \hline
			\rowcolor[HTML]{C6D8F0} 
			initial\_max\_stream\_data\_uni           & QUIC                        & QUIC parameters     & $q_7$                       & numerical               & low                     \\ \hline
			\rowcolor[HTML]{C6D8F0} 
			initial\_max\_streams\_bidi               & QUIC                        & QUIC parameters     & $q_8$                       & numerical               & low                     \\ \hline
			\rowcolor[HTML]{C6D8F0} 
			initial\_max\_streams\_uni                & QUIC                        & QUIC parameters     & $q_9$                       & numerical               & low                     \\ \hline
			\rowcolor[HTML]{C6D8F0} 
			max\_ack\_delay                           & QUIC                        & QUIC parameters     & $q_{10}$                      & numerical               & low                     \\ \hline
			\rowcolor[HTML]{C6D8F0} 
			disable\_active\_migration                & QUIC                        & QUIC parameters     & $q_{11}$                      & presence                & low                     \\ \hline
			\rowcolor[HTML]{C6D8F0} 
			active\_connection\_id\_limit             & QUIC                        & QUIC parameters     & $q_{12}$                      & numerical               & low                     \\ \hline
			\rowcolor[HTML]{C6D8F0} 
			initial\_source\_connection\_id           & QUIC                        & QUIC parameters     & $q_{13}$                      & length                  & low                     \\ \hline
			\rowcolor[HTML]{C6D8F0} 
			max\_datagram\_frame\_size                & QUIC                        & QUIC parameters     & $q_{14}$                      & numerical               & low                     \\ \hline
			\rowcolor[HTML]{C6D8F0} 
			grease\_quic\_bit                         & QUIC                        & QUIC parameters     & $q_{15}$                      & presence                & low                     \\ \hline
			\rowcolor[HTML]{C6D8F0} 
			initial\_rtt                              & QUIC                        & QUIC parameters     & $q_{16}$                      & presence                & low                     \\ \hline
			\rowcolor[HTML]{C6D8F0} 
			google\_connection\_options               & QUIC                        & QUIC parameters     & $q_{17}$                      & categorical             & medium                  \\ \hline
			\rowcolor[HTML]{C6D8F0} 
			user\_agent                               & QUIC                        & QUIC parameters     & $q_{18}$                      & categorical             & medium                  \\ \hline
			\rowcolor[HTML]{C6D8F0} 
			google\_version                           & QUIC                        & QUIC parameters     & $q_{19}$                      & categorical             & medium                  \\ \hline
			\rowcolor[HTML]{C6D8F0} 
			version\_information                      & QUIC                        & QUIC parameters     & $q_{20}$                      & categorical             & medium                  \\ \hline
		\end{tabular}\label{tab:AppendixFeatureList}
	}
\end{table*}

\subsubsection{Creating attributes from fields contained in handshake messages}\label{sec:classification-attributes-cost}
For the purpose of feeding the fields of interest into our machine learning models, we seek to first convert them into numerical attributes. We begin by considering those that are inherently numerical, such as \textit{handshake\_length} and \textit{extensions\_length}, for which no transformation is needed.

Fields that are categorical take values from a finite set of elements. For example, \textit{compress\_certificate} can take one of 
\textit{zlib} and \textit{brotli}. We assign a unique positive integer to them, \ie 1 and 2 respectively, and transform this field into a numerical attribute. We assign 0 to categorical fields that are not present in a flow.

Fields that are lists can in turn include many categorical fields. For example, \textit{cipher\_suites} contains multiple cipher suites from a finite list supported by a client. The order of the categorical items in the field indicates the client's preference. Therefore, to preserve the information provided by the choice of items and their order in a list-type field, we use a fixed-length vector (\ie by encoding the list as separate positional attributes) to indicate the placement of each item, with zero-padding for non-existing items. This simplified representation can be readily processed by the classifiers.

For 17 fields such as \textit{grease\_quic\_bit}, they do not have any associated value. However, whether they are present or not differs across device types and software agents. Therefore, we assign 1 for their presence in a video flow and 0 otherwise. Also, for 7 fields such as \textit{initial\_source\_connection\_id} in QUIC, the values they contain are not useful since they are randomly chosen \cite{rfc9000}, but the length of the values, in terms of the number of bytes they contain, could be helpful, and hence we treat them as length-based attributes. 

\begin{figure*}[t]
	\mbox{
		\hspace{-3mm}
		\subfigure[Attribute importance for \textbf{\color{alizarin}YouTube} \textbf{QUIC} flows.]{
			{\includegraphics[width=1.08\columnwidth]{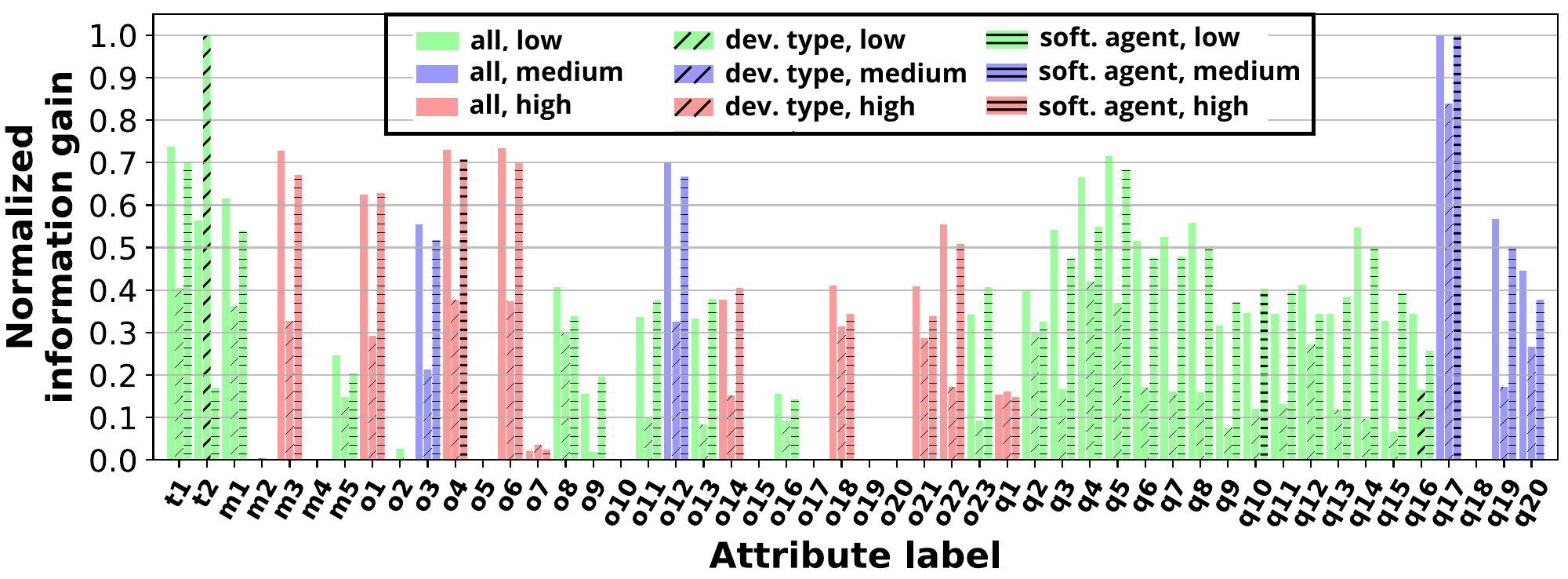}
				\label{fig:feature_importance_quic}}
		}
		\hspace{-2mm}
		\subfigure[Attribute importance for \textbf{\color{alizarin}YouTube} \textbf{TCP} flows.]{
			{\includegraphics[width=0.95\columnwidth]{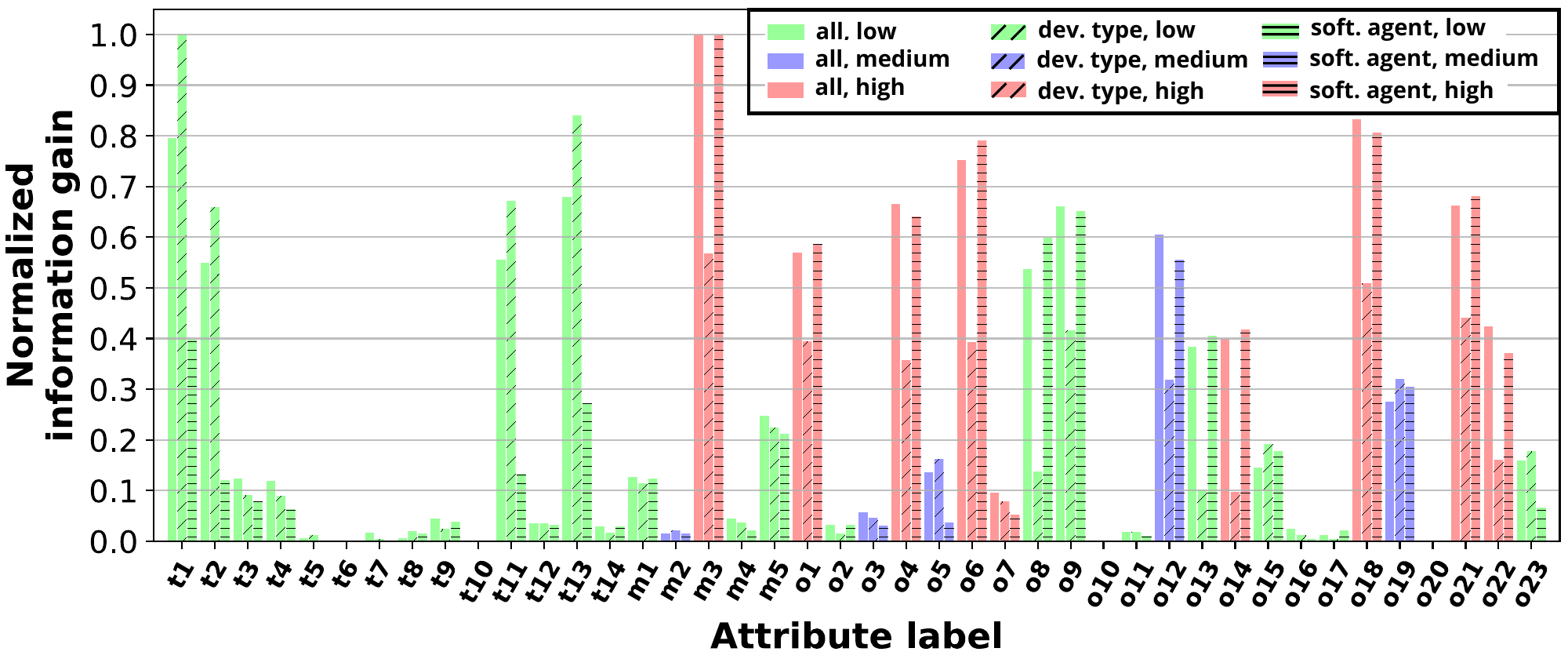}
				\label{fig:feature_importance_tcp}}
		}
	}
	\vspace{-5mm}
	\caption{Attribute importance in different classification objectives including user platforms, only device types, or only software agents for YouTube (a) QUIC and (b) TCP flows. The level of preprocessing required (annotated by low-, medium- or high-cost) and classification objectives are represented by different colors and patterns, respectively.}
	\label{fig:feature_importance}
\end{figure*}

We note that converting each field to its corresponding attribute (\ie the green box in Fig.~\ref{fig:pipeline}) can come with different levels of processing steps that can require different levels of computational costs, especially when deployed for high-speed broadband networks in real time.
In our real-time packet processing pipeline as shown in Fig.~\ref{fig:pipeline}, numerical fields are directly taken from handshake packets as input attributes. Length-based and presence-based attributes are also directly taken from their corresponding header fields without additional processing. These attributes can hence be considered low-cost.
Attributes that are converted from categorical fields each requires an additional value mapping process after packet parsing, which is typically done through basic dictionary lookup operations. Albeit small in terms of time consumption (\eg several microseconds), there is an additional processing step involved, hence these attributes are considered medium-cost. 
A list-type field can contain multiple (categorical) items, each requiring an additional value mapping process to convert the entire list-type field into one attribute in the form of an array of numerical values. For example, the list-type field \textit{cipher\_suites} often has over ten \textit{cipher\_suite} items. To convert one such field into its corresponding attribute (\ie $m_3$) as an array of numerical values, the system needs to loop through every single item in the list, resulting in a total of over ten additional value mapping processes required. Therefore, such attributes are considered high-cost.
As we will show next, using attributes with higher preprocessing costs in classification tasks does not necessarily increase the overall predictive capability.

\subsubsection{Importance of attributes}	\label{sec:attribute-importance}
As discussed in \S\ref{sec:analysis-handshake}, the attributes derived from the handshake fields are not all equally important in predicting user platforms. We now systematically benchmark the importance of each attribute using the \emph{information gain} metric \cite{quinlan_induction_1986}. 
It is indeed the mutual information between an attribute and the prediction target, which is calculated as the sum of the entropies of the attribute and the predicted class deduced by their joint entropy \cite{kozachenko_sample_1987,sklearn_mutual_nodate}.
Therefore, attributes of the highest importance have information gain close to 1 and an irrelevant attribute has information gain 0.
In our analysis, the information gains are computed for both TCP and QUIC video flows for each of the four video providers, with prediction objectives being user platform, device type, and software agent. We now discuss the relative importance of the attributes using YouTube video flows over QUIC as a representative example. 

Fig.~\ref{fig:feature_importance_quic} shows the importance (\ie normalized information gain) of attributes for YouTube over QUIC flows. Each bar is color-coded by its level of preprocessing required and pattern-coded by its prediction objective. Attributes are denoted by their ordered labels, and a full mapping between labels and attribute names is provided in Table~\ref{tab:AppendixFeatureList}.

To elicit the relative importance of attributes in satisfying our prediction objectives, we empirically define two thresholds, 0.2 and 0.1, and rate the importance of attributes as high, medium or low if their information gain value is $>0.2$, between $0.1$ and $0.2$ or $<0.1$, respectively. We can see that 17 attributes including $t_{1}$, $m_{1}$, $m_{3}$, $o_{1}$, $o_{3}$, $o_{4}$, $o_{6}$, $o_{8}$, $o_{12}$, $o_{18}$, $o_{21}$, $q_{2}$, $q_{4}$, $q_{5}$, $q_{12}$, $q_{17}$ and $q_{20}$ have high importance for all three prediction objectives, \ie user platform, only device type, and only software agent.
Eleven attributes, \ie $m_{2}$, $m_{4}$, $o_{2}$, $o_{5}$, $o_{7}$, $o_{10}$, $o_{15}$, $o_{17}$, $o_{19}$, $o_{20}$ and $q_{18}$, have information gain $<0.1$ for all three prediction objectives, implying their effectiveness is limited.

Other attributes have at least one high or medium score, and one medium or low score. For instance, $t_{2}$ has a normalized importance score of 1 (high) for device type but 0.18 (medium) for software agent. This is not surprising since certain handshake fields (\eg time-to-live) are highly dependent on device types while others (\eg version information) exhibit strong variations across software agents; see \S\ref{sec:analysis-handshake}. In addition, as shown in Fig.~\ref{fig:feature_importance_tcp}, an attribute with low importance for QUIC video flows may have medium or high importance for TCP flows. An example is $o_{15}$, which has a value near 0 for QUIC but over 0.1 for TCP. These observations indicate that a select combination of attributes with medium to high importance can also serve as useful predictors.

Finally, we highlight that even though certain types of attributes require only zero or minimal preprocessing steps, such as the numerical fields in handshake messages, relying solely on them does not necessarily compromise prediction accuracy. For instance, out of 43 numerical and length-/presence-based (\ie low-cost) attributes, 3 (\ie $t_{1}$, $t_{2}$, $o_{8}$) have high importance for both QUIC and TCP video flows. On the other hand, 4 out of 9 categorical (\ie medium-cost) and 1 out of 10 list-type (\ie high-cost) attributes have low importance. In production environments, switches might not have the necessary computational resources to execute the entire packet processing and classification pipeline in real-time. Under these circumstances, one can (carefully) select only those attributes requiring minimal preprocessing to achieve a certain acceptable degree of prediction accuracy. This trade-off between the cost of preprocessing attributes and the accuracy of predicting user platforms is discussed next.

\subsection{Classification Models}\label{sec:classification-models}
We now present our machine learning models to predict user platforms for the four video content providers. Note that at the time of this writing, only YouTube supports QUIC while the others support only TCP. 

\begin{figure}[!t]
	\mbox{
		\hspace{-3mm}
		\subfigure[Tuning Random Forest model.]{
			{\includegraphics[width=0.45\linewidth]{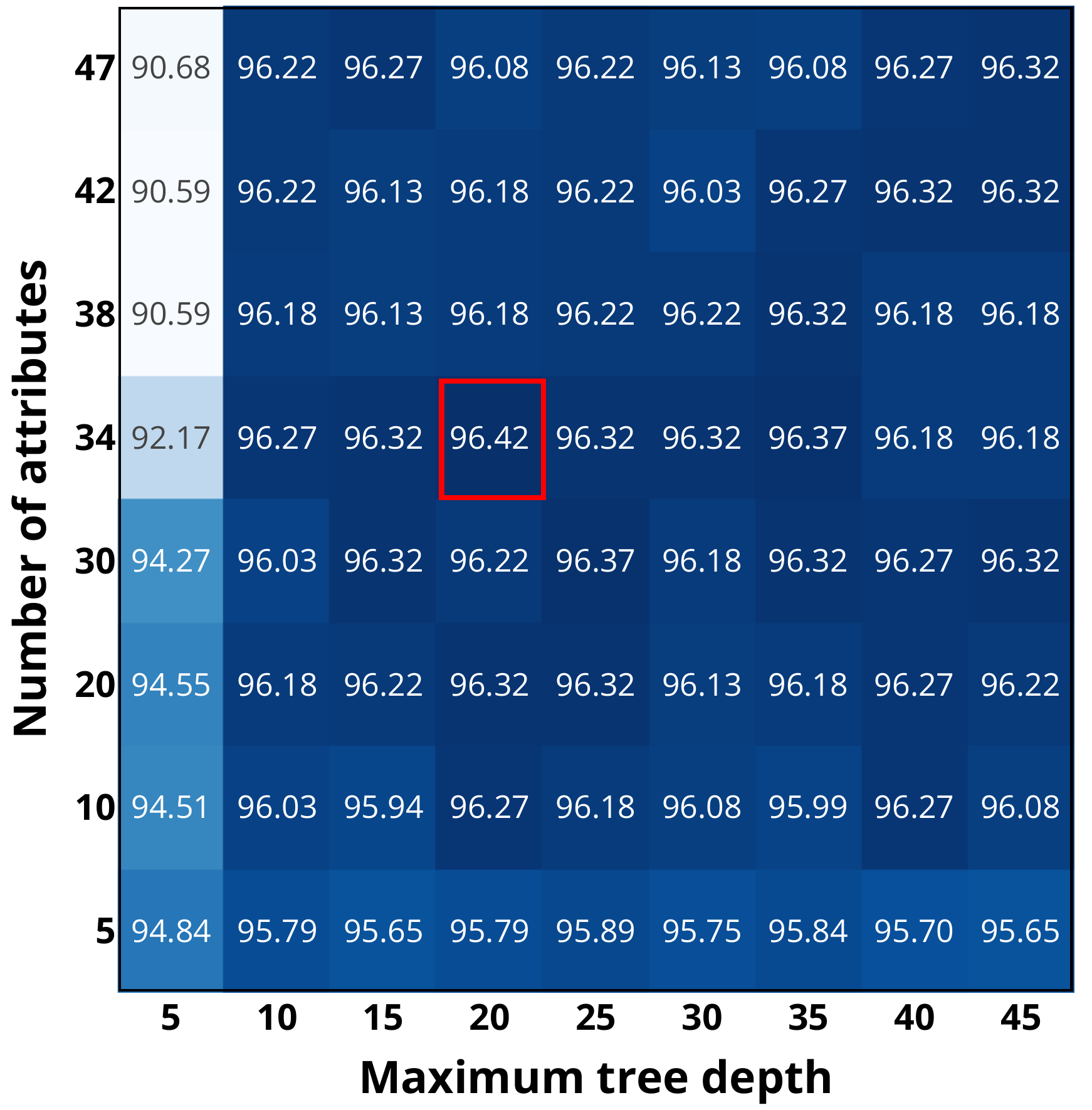}}\quad
			\label{fig:model_tuning}
		}
		\hspace{-5mm}
		\subfigure[Accuracy for user platform.]{
			{\includegraphics[width=0.53\linewidth]{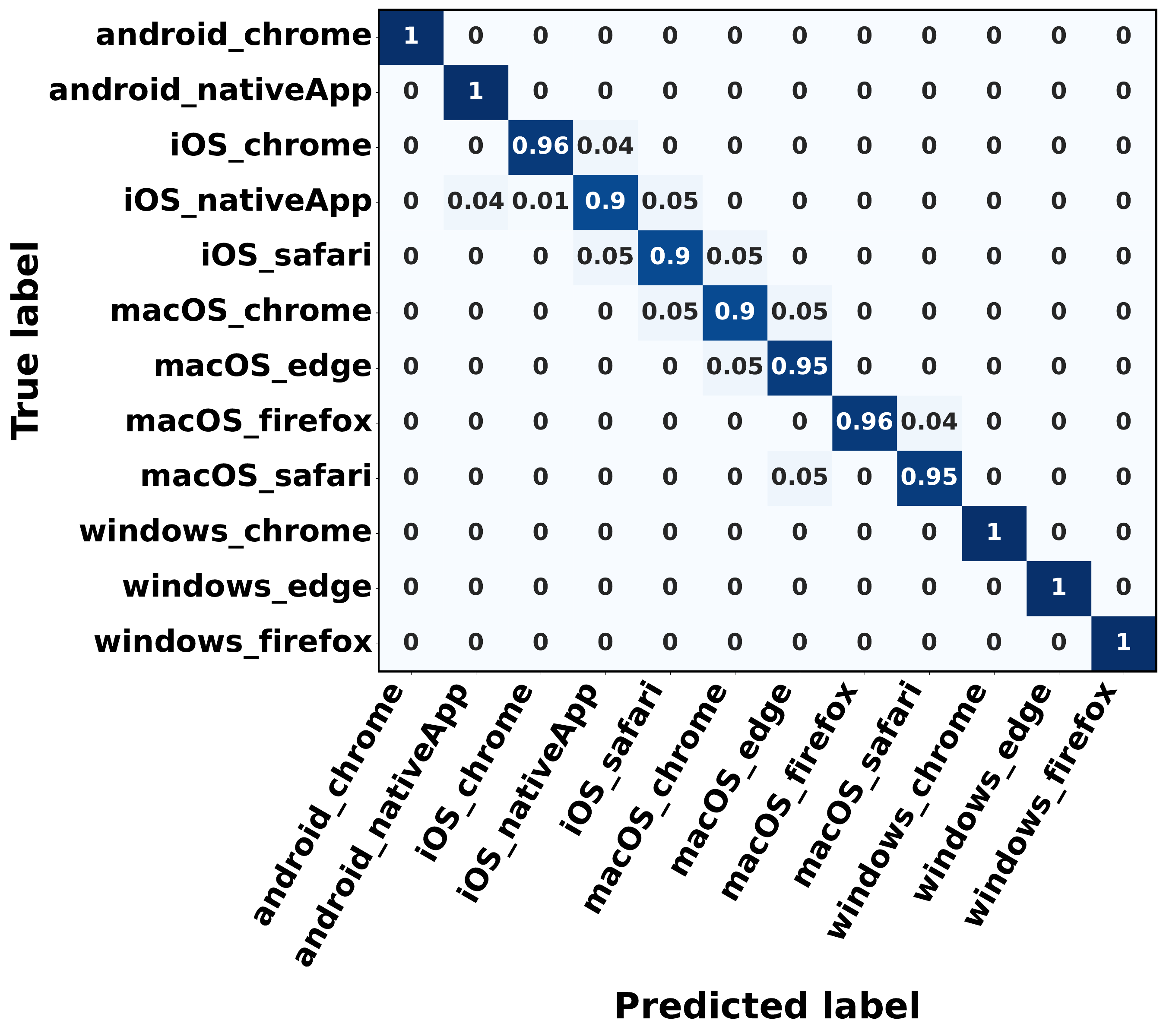}}\quad
			\label{fig:best_model}
		}								
	}

	\mbox{
		\hspace{-3mm}
		\subfigure[Accuracy for only device type.]{
			{\includegraphics[width=0.48\linewidth]{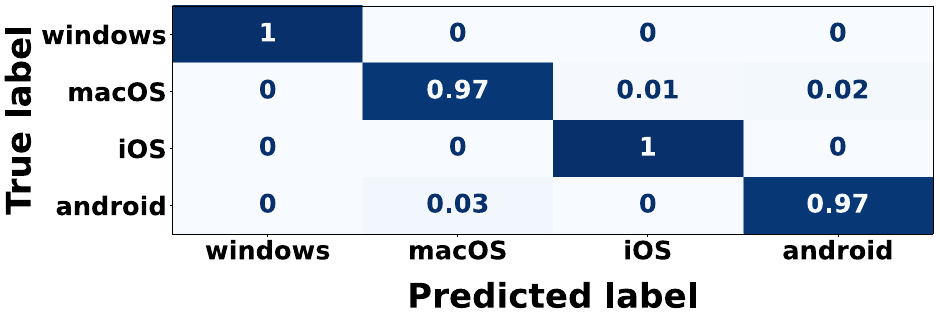}}\quad
			\label{fig:best_model-OS}
		}
		\hspace{-4mm}
		\subfigure[Accuracy for only software agent.]{
			{\includegraphics[width=0.48\linewidth]{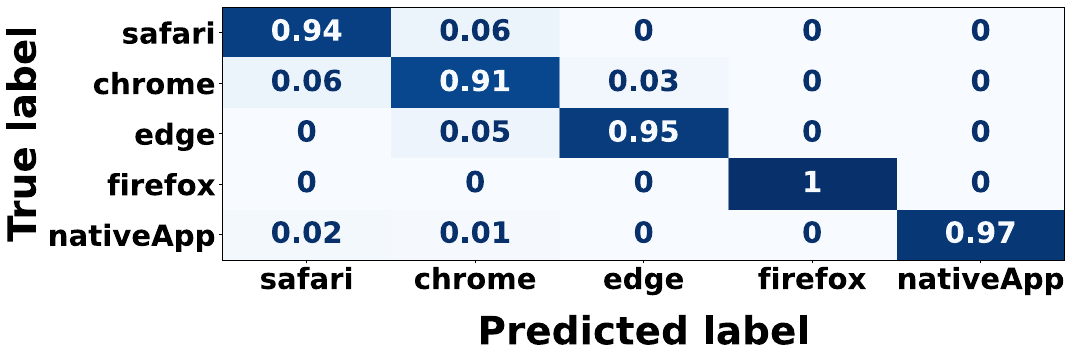}}\quad
			\label{fig:best_model-ST}
		}								
	} 
	\vspace{-3mm}
	\caption{Hyperparameter tuning of random forest models for YouTube over QUIC (a). The model's classification accuracy to classify user platform is shown in (b), only device type is depicted in (c), and only software agent is presented in (d).}
	\label{fig:confusion_matrix}
\end{figure}

\subsubsection{Model training, tuning and selection}
We consider three popular machine learning algorithms for our classification tasks, \ie random forest (decision tree-based), MLP (neural network), and KNN (clustering-based), and tune their hyperparameters accordingly. For random forest, these parameters include maximum tree depth, number of trees, and number of attributes. MLP models are tuned for the number of hidden layers, number of perceptrons per layer and activation functions. KNN classifiers are tuned for the number of neighbors, weight functions and leaf size. 

We trained each of these algorithms on the dataset collected from our experimental setup shown in Fig.~\ref{fig:lab_setup}. To recall, it consists of attributes extracted from traces spanning about 10,000 video flows across 30 user platforms. The performance of each of the models was evaluated in terms of the overall accuracy using 10-fold cross-validation. Our evaluation shows that the random forest model outperforms both the MLP and KNN classifiers, not only for YouTube but for all other providers and user platforms as well. For example, in classifying the user platform for YouTube flows over QUIC, the random forest model achieves an overall accuracy of 96.4\%, while the MLP and KNN models achieve an accuracy of 65.1\% and 69.1\%, respectively. This observation is consistent with prior research that shows decision tree-based models are better suited for network traffic classification problems \cite{resende_2018_survey}. We therefore selected the random forest model for real-time deployment and evaluation in our campus network. 

Fig.~\ref{fig:model_tuning} depicts how we select the best random forest model. Out of the 62 attributes overall, only 50 are applicable to QUIC. Further, the figure shows the overall classification accuracy when tuning the number of attributes (vertical axis) and the maximum tree depth (horizontal axis). The highest test accuracy of 96.4\% was attained when the two hyperparameters were set to 34 and 20, respectively. We use this selection of hyperparameters as our best-performing random forest model. 

For the above model, we show its classification accuracy per prediction class (\ie confusion matrix) in Fig.~\ref{fig:best_model}. It is clear that 5 out of 12 prediction classes are with 100\% accuracy, including all browser types on Windows PC, as well as Chrome and native YouTube application on Android phone. Misclassified instances are observed within two groups, \ie iOS and macOS devices. For example, native YouTube app on iOS has a small chance (\ie $\leq$ 4\%) of being misclassified as native app on Android. For the iOS native app instances, their device types (\ie iOS) are classified with 96\% accuracy, while their software agents (\ie native app) might be misclassified as Chrome or Safari with a slim chance of less than 6\%. 

Delving deeper into the fields for (iOS, Safari) and (iOS, Chrome) platforms, see columns 10 and 11 in Fig.~\ref{fig:heatmap_quic}, we note that a vast majority of them take on similar values. Only a small number of attributes, \eg \textit{handshake\_length}, \textit{extensions\_length} have different values. Moreover, Chrome has started to randomize TLS extension orders since version 110 \cite{a2023_chrome}. These variations in attribute values explain why we see a small fraction of misclassifications in these user platforms.

We also observe that those misclassified instances are with low confidence, \ie less than 50\%, while the correctly classified ones are with high confidence, \ie over 80\%. The performance of our random forest models in classifying only the device type and the software agent for YouTube QUIC is depicted in Fig~\ref{fig:best_model-OS} and \ref{fig:best_model-ST}. We note that their accuracy in identifying the device type is high, $\ge$ 97\% for all device types, while it can predict all software agents with $\ge$ 91\% accuracy. The marginal decline in the accuracy of the latter is due to the reasons explained above. Nonetheless, the overall results demonstrate that our random forest classifiers offer high prediction accuracy with high confidence.

\begin{table}[t]
	\caption{Model performance in open-set evaluation for three classification objectives, namely user platform, device type only and software agent only.}
	\vspace{-3mm}
        \small
	\begin{tabular}{|
		>{\columncolor[HTML]{EFEFEF}}c |l|l|}
		\hline
		\textbf{Provider}                                                                                       & \textbf{Objective} & \textbf{Accuracy} \\ \hline
		\cellcolor[HTML]{EFEFEF}                                                                                & User platform      & 98.7\%/94.5\%     \\ \cline{2-3} 
		\cellcolor[HTML]{EFEFEF}                                                                                & Device type        & 99.1\%/98.4\%     \\ \cline{2-3} 
		\multirow{-3}{*}{\cellcolor[HTML]{EFEFEF}\begin{tabular}[c]{@{}c@{}}YouTube \\ (TCP/QUIC)\end{tabular}} & Software agent     & 96.6\%/95.4\%     \\ \hline
		\cellcolor[HTML]{EFEFEF}                                                                                & User platform      & 91.2\%            \\ \cline{2-3} 
		\cellcolor[HTML]{EFEFEF}                                                                                & Device type        & 92.4\%            \\ \cline{2-3} 
		\multirow{-3}{*}{\cellcolor[HTML]{EFEFEF}\begin{tabular}[c]{@{}c@{}}Netflix\\ (TCP)\end{tabular}}       & Software agent     & 90.6\%            \\ \hline
		\cellcolor[HTML]{EFEFEF}                                                                                & User platform      & 90.9\%            \\ \cline{2-3} 
		\cellcolor[HTML]{EFEFEF}                                                                                & Device type        & 91.6\%            \\ \cline{2-3} 
		\multirow{-3}{*}{\cellcolor[HTML]{EFEFEF}\begin{tabular}[c]{@{}c@{}}Disney\\ (TCP)\end{tabular}}        & Software agent     & 88.6\%            \\ \hline
		\cellcolor[HTML]{EFEFEF}                                                                                & User platform      & 88.2\%            \\ \cline{2-3} 
		\cellcolor[HTML]{EFEFEF}                                                                                & Device type        & 89.4\%            \\ \cline{2-3} 
		\multirow{-3}{*}{\cellcolor[HTML]{EFEFEF}\begin{tabular}[c]{@{}c@{}}Amazon\\ (TCP)\end{tabular}}        & Software agent     & 87.9\%            \\ \hline
		\end{tabular}
	\label{tab:openset}
\end{table}

\begin{table}[t]
	\caption{Median classification confidence of correct and incorrect instances in the open-set evaluation for all four providers and three classification objectives.}
	\vspace{-3mm}
	\small
	\begin{tabular}{|
			>{\columncolor[HTML]{EFEFEF}}l |l|l|l|}
		\hline
		\textbf{Provider}                                                                                      & \textbf{Objective} & \textbf{\begin{tabular}[c]{@{}l@{}}Med. conf. \\ (correct)\end{tabular}} & \textbf{\begin{tabular}[c]{@{}l@{}}Med. conf. \\ (incorrect)\end{tabular}} \\ \hline
		\cellcolor[HTML]{EFEFEF}                                                                               & User platform      & 98.5\%/91.4\%                                                             & 86.5\%/54.4\%                                                               \\ \cline{2-4} 
		\cellcolor[HTML]{EFEFEF}                                                                               & Device type        & 89.6\%/91.8\%                                                             & 46.7\%/57.5\%                                                               \\ \cline{2-4} 
		\multirow{-3}{*}{\cellcolor[HTML]{EFEFEF}\begin{tabular}[c]{@{}l@{}}YouTube\\ (TCP/QUIC)\end{tabular}} & Software agent     & 98.2\%/90.9\%                                                             & 89.3\%/52.7\%                                                               \\ \hline
		\cellcolor[HTML]{EFEFEF}                                                                               & User platform      & 88.7\%                                                                    & 53.9\%                                                                      \\ \cline{2-4} 
		\cellcolor[HTML]{EFEFEF}                                                                               & Device type        & 99.3\%                                                                    & 60.0\%                                                                      \\ \cline{2-4} 
		\multirow{-3}{*}{\cellcolor[HTML]{EFEFEF}\begin{tabular}[c]{@{}l@{}}Netflix\\ (TCP)\end{tabular}}      & Software agent     & 91.0\%                                                                    & 59.1\%                                                                      \\ \hline
		\cellcolor[HTML]{EFEFEF}                                                                               & User platform      & 91.5\%                                                                    & 67.6\%                                                                      \\ \cline{2-4} 
		\cellcolor[HTML]{EFEFEF}                                                                               & Device type        & 98.2\%                                                                    & 83.5\%                                                                      \\ \cline{2-4} 
		\multirow{-3}{*}{\cellcolor[HTML]{EFEFEF}\begin{tabular}[c]{@{}l@{}}Disney\\ (TCP)\end{tabular}}       & Software agent     & 91.6\%                                                                    & 67.6\%                                                                      \\ \hline
		\cellcolor[HTML]{EFEFEF}                                                                               & User platform      & 89.1\%                                                                    & 60.6\%                                                                      \\ \cline{2-4} 
		\cellcolor[HTML]{EFEFEF}                                                                               & Device type        & 99.4\%                                                                    & 50.0\%                                                                      \\ \cline{2-4} 
		\multirow{-3}{*}{\cellcolor[HTML]{EFEFEF}\begin{tabular}[c]{@{}l@{}}Amazon\\ (TCP)\end{tabular}}       & Software agent     & 91.3\%                                                                    & 64.3\%                                                                      \\ \hline
	\end{tabular}
	\label{tab:openset-confidence}
\end{table}

\subsubsection{Open-set evaluation}
Relying solely on the accuracy reported by 10-fold validation can result in over-fitting, which can mask the true accuracy of a classifier. 
To overcome this problem, we further evaluate the performance of our random forest model on a dataset collected by one of the authors from their home network. While the devices in the home are the same as those in our experimental setup, the OS versions as well as those of the software agents are different. These variations could impact the values of the different attributes. The aim of this exercise therefore is to validate the accuracy of the model, trained on the lab trace data, in predicting the user platforms seen in a different environment, \ie the home.

This dataset contains over 2000 video flows spread evenly across all user platforms. We note from Table~\ref{tab:openset} that the results are comparable to the ones reported earlier, confirming the high efficacy of the classifier. User platforms are classified with $>94\%$ accuracy for YouTube, $>91\%$ for Netflix, $>90\%$ for Disney+ and $>88\%$ for Amazon Prime Video. 
Additionally, as shown in Table~\ref{tab:openset-confidence}, the majority of correctly classified instances are of very high confidence (\ie $>88\%$) whereas incorrect classifications tend to be of low confidence (\ie $<70\%$). Exceptions are also observed in minor cases for certain types of classes due to their similar networking and kernel suites to the misclassified classes. For example, video flows from Apple's mobile iOS devices sometimes behave very similarly to Apple's desktop macOS devices, thus, can be misclassified with high confidence. 

\subsubsection{Models with a subset of attributes}
We have discussed in \S\ref{sec:attribute-importance} that not all attributes are equally important from the perspective of information gain, especially list attributes which require high costs for preprocessing. An ISP carrying very high data rates, \eg multi-hundreds of Gbps, may require servers with significant computational capability to deploy the end-to-end classification pipeline in real-time. In the absence of such resources, one may choose to discard the use of high-cost low-importance attributes to reduce the processing load with negligible impact on classification accuracy, as described next. 

We train random forest models with three subsets of attributes. Each subset excludes attributes that are deemed to be of low importance (\ie $<0.1$ information gain). Additionally, the first subset excludes only high-cost low-importance attributes, the second subset excludes both high- and medium-cost attributes that are of low importance, and the third subset excludes all attributes that are of low importance.

The overall accuracy of the random forest classifier for YouTube QUIC flows is reported in  Table~\ref{tab:subset-attribute}. Compared with the accuracy of the model that uses the full (\ie 50) attribute set, we observe a slight reduction ($\approx$ 3\%) across the three scenarios considered here. For example, the accuracy for classifying user platforms of QUIC YouTube flows is 96.4\% with the full attribute set. This drops to 93.3\%, 93.0\% and 92.8\%, with each subset, respectively. 
The performance is similar when predicting only the device type or software agent. 

\begin{table}[t]
	\caption{Accuracy of models for YouTube QUIC video flows with three subsets of attributes. Each subset excludes low-importance attributes associated with high cost; high and medium cost; high, medium and low cost.}
	\vspace{-3mm}
        \small
	\begin{tabular}{|l|l|c|}
	\hline
	\rowcolor[HTML]{EFEFEF} 
	\textbf{{\color{darkred}Excluded} low-imp. attributes}                         & \textbf{Objective} & \textbf{Accuracy} \\ \hline
														 & User platform                    & 93.3\%            \\ \cline{2-3} 
														 & Device type                       & 97.2\%            \\ \cline{2-3} 
	\multirow{-3}{*}{\textbf{High} cost}                 & Software agent                 & 94.6\%            \\ \hline
														 & User platform                    & 93.0\%            \\ \cline{2-3} 
														 & Device type                       & 97.2\%            \\ \cline{2-3} 
	\multirow{-3}{*}{\textbf{High} + \textbf{medium} cost}        & Software agent                 & 92.8\%            \\ \hline
														 & User platform                    & 92.8\%            \\ \cline{2-3} 
														 & Device type                       & 97.1\%            \\ \cline{2-3} 
	\multirow{-3}{*}{\textbf{High} + \textbf{medium} + \textbf{low} cost} & Software agent                 & 92.9\%            \\ \hline
	\end{tabular}
	\label{tab:subset-attribute}
	\end{table}

Thus, in scenarios with constrained computational resources, one can safely ignore attributes with low information gain requiring varying degrees of processing costs, to deploy the end-to-end user platform classification pipeline in real-time. 
In our campus-wide deployment, our server has sufficient computational resources to handle real-time traffic streams at 20 Gbps peak rate and process the observed maximum of over 1000 concurrent video flows from considered content providers. Thus, we use classifiers trained with the full attribute set to achieve the highest possible accuracy.

\begin{table*}[t]
	\caption{Benchmarking the classification accuracy of our user platform identification method against state-of-the-art after necessary methodological adaptations.}
	\vspace{-3mm}
	\fontsize{7pt}{8.5pt}\selectfont
	\begin{tabular}{|l|l|
		>{\columncolor[HTML]{FFFFC7}}l |l|l|l
		>{\columncolor[HTML]{9AFF99}}l lll|}
		\hline
		\cellcolor[HTML]{EFEFEF}                         & \cellcolor[HTML]{EFEFEF}                             & \cellcolor[HTML]{EFEFEF}                                       & \cellcolor[HTML]{EFEFEF}                                        & \cellcolor[HTML]{EFEFEF}                                             & \multicolumn{5}{l|}{\cellcolor[HTML]{EFEFEF}\textbf{Accuracy after adaptation}}                                                                                                                                                                                                                   \\ \cline{6-10} 
		\multirow{-2}{*}{\cellcolor[HTML]{EFEFEF}\textbf{Work}} & \multirow{-2}{*}{\cellcolor[HTML]{EFEFEF}\textbf{Objective}} & \multirow{-2}{*}{\cellcolor[HTML]{EFEFEF}\textbf{Protocol}} & \multirow{-2}{*}{\cellcolor[HTML]{EFEFEF}\textbf{Granularity}} & \multirow{-2}{*}{\cellcolor[HTML]{EFEFEF}\textbf{Required specific adaptations}} & \multicolumn{1}{l|}{\cellcolor[HTML]{EFEFEF}\textbf{{\color{alizarin}YT} (QUIC)}} & \multicolumn{1}{l|}{\cellcolor[HTML]{EFEFEF}\textbf{{\color{alizarin}YT} (TCP)}} & \multicolumn{1}{l|}{\cellcolor[HTML]{EFEFEF}\textbf{{\color{blue-violet}NF} (TCP)}} & \multicolumn{1}{l|}{\cellcolor[HTML]{EFEFEF}\textbf{{\color{blizzardblue}DN} (TCP)}} & \cellcolor[HTML]{EFEFEF}\textbf{{\color{blue}AP} (TCP)} \\ \hline
		Ours                                             & \cellcolor[HTML]{9AFF99}Dev. type + Soft. agent & \cellcolor[HTML]{9AFF99}TLS/QUIC                             & \cellcolor[HTML]{9AFF99}flow                                    & ---                                                                  & \multicolumn{1}{l|}{\cellcolor[HTML]{9AFF99}94.5\%}         & \multicolumn{1}{l|}{\cellcolor[HTML]{9AFF99}98.7\%}        & \multicolumn{1}{l|}{\cellcolor[HTML]{9AFF99}91.2\%}        & \multicolumn{1}{l|}{\cellcolor[HTML]{9AFF99}90.9\%}       & \cellcolor[HTML]{9AFF99}88.2\%       \\ \hline
		\cite{anderson_tls_2019}                        & \cellcolor[HTML]{9AFF99}Dev. type + Soft. agent & TLS                                                            & \cellcolor[HTML]{9AFF99}flow                                    & \cellcolor[HTML]{FFCCC9}feature construct.; classi. process & \multicolumn{1}{l|}{\cellcolor[HTML]{9AFF99}90.1\%}         & \multicolumn{1}{l|}{\cellcolor[HTML]{9AFF99}97.5\%}        & \multicolumn{1}{l|}{\cellcolor[HTML]{FFFFC7}84.0\%}        & \multicolumn{1}{l|}{\cellcolor[HTML]{FFFFC7}82.8\%}       & \cellcolor[HTML]{FFFFC7}80.3\%       \\ \hline
		\cite{fan_identify_2019}                           & \cellcolor[HTML]{FFFFC7}Dev. type                  & TLS                                                            & \cellcolor[HTML]{FFFFC7}host                                    & \cellcolor[HTML]{FFCCC9}flow granularity; inference object.        & \multicolumn{1}{l|}{\cellcolor[HTML]{9AFF99}94.0\%}         & \multicolumn{1}{l|}{\cellcolor[HTML]{9AFF99}96.8\%}        & \multicolumn{1}{l|}{\cellcolor[HTML]{FFFFC7}86.0\%}        & \multicolumn{1}{l|}{\cellcolor[HTML]{FFFFC7}80.1\%}       & \cellcolor[HTML]{9AFF99}84.1\%       \\ \hline
		\cite{lastovicka_using_2020}                     & \cellcolor[HTML]{FFFFC7}Dev. type                  & TLS                                                            & \cellcolor[HTML]{FFFFC7}host                                    & \cellcolor[HTML]{FFCCC9}flow granularity; inference object.       & \multicolumn{1}{l|}{\cellcolor[HTML]{FFCCC9}68.1\%}         & \multicolumn{1}{l|}{\cellcolor[HTML]{9AFF99}95.1\%}        & \multicolumn{1}{l|}{\cellcolor[HTML]{FFFFC7}82.7\%}        & \multicolumn{1}{l|}{\cellcolor[HTML]{FFFFC7}83.1\%}       & \cellcolor[HTML]{FFFFC7}79.0\%       \\ \hline
		\cite{richardson_Novel_2020}          & \cellcolor[HTML]{9AFF99}Dev. type + Soft. agent              & \cellcolor[HTML]{FFCCC9} non-TLS                           & \cellcolor[HTML]{FFFFC7}host                                   & \cellcolor[HTML]{FD6864}not adaptable                                  & \multicolumn{1}{l|}{---}                                        & \multicolumn{1}{l|}{---}                                       & \multicolumn{1}{l|}{---}                                       & \multicolumn{1}{l|}{---}                                       & ---                                       \\ \hline
		\cite{ren_app_2021}               & \cellcolor[HTML]{FFFFC7}Soft. agent               & TLS                                                            & \cellcolor[HTML]{9AFF99}flow                                    & \cellcolor[HTML]{FFFFC7}inference objective                          & \multicolumn{1}{l|}{\cellcolor[HTML]{FD6864}11.3\%}         & \multicolumn{1}{l|}{\cellcolor[HTML]{FFCCC9}51.0\%}        & \multicolumn{1}{l|}{\cellcolor[HTML]{FFCCC9}53.4\%}        & \multicolumn{1}{l|}{\cellcolor[HTML]{FFCCC9}56.5\%}       & \cellcolor[HTML]{FFCCC9}38.1\%       \\ \hline
		\cite{marzani_Mobile_2023}         & \cellcolor[HTML]{FFFFC7}Soft. agent                          & TLS                                                          & \cellcolor[HTML]{FFFFC7}host                                   & \cellcolor[HTML]{FD6864}not adaptable                                  & \multicolumn{1}{l|}{---}                                        & \multicolumn{1}{l|}{---}                                       & \multicolumn{1}{l|}{---}                                       & \multicolumn{1}{l|}{---}                                       & ---                                       \\ \hline
	\end{tabular}
	\label{tab:compare_sota}
\end{table*}

\subsubsection{Benchmarking against state-of-the-art}
\label{sec:benchmark}
Now, we benchmark our method with six state-of-the-art techniques spanning the last five years of literature in user platform identification using the ground-truth dataset discussed in \S\ref{sec:analysis-dataset}. Three qualitative aspects, as specified in the second to fourth columns of Table~\ref{tab:compare_sota}, including inference objective, covered protocol and inference granularity are compared to demonstrate the superior visibility our method can provide. 
Our method outperforms all alternatives.

Two out of the six prior techniques \cite{richardson_Novel_2020,marzani_Mobile_2023} require collecting statistics of all flows from a candidate host, and thus cannot be used to identify user platforms of individual video flows from clients behind NAT.
The other four techniques \cite{anderson_tls_2019,fan_identify_2019,lastovicka_using_2020,ren_app_2021}, either directly offer flow-level granularity, or have a subset of articulated attributes from individual flows, and thus can be adapted for flow-level identification as specified in the fifth column of Table~\ref{tab:compare_sota}. 
First of all, since all six prior techniques are designed only for TLS over TCP flows, a generic adaptation has to be made for all prior works to handle video flows over QUIC, including identifying and decrypting QUIC Initial packets and extracting handshake attributes from TLS CHLO messages over QUIC.
In addition, to make the four techniques applicable for user platform identification, the required adaptations specific to each prior work are listed in the fifth column of Table~\ref{tab:compare_sota} such as extracting fine-grained flow-level telemetry, constructing features from collected statistics, expanding inference objective, developing classification pipeline and models. For example, the method in \cite{anderson_tls_2019} combines TLS handshake fields to generate string-based fingerprints of applications. Therefore, for benchmarking purposes, we adapt this method by constructing usable features from their fingerprint strings and developing a classification process.
Also, the works in \cite{fan_identify_2019} and \cite{lastovicka_using_2020} classify device types with IP-level attributes. Thus, their attributes are adapted to be extracted for individual video flows and used to classify not only device types but also software agents.

The quantitative accuracy of our method in comparison to other prior techniques (after necessary adaptations) is reported in the last five columns of Table~\ref{tab:compare_sota}. We can clearly observe that our method outperforms other techniques in all five classification scenarios, \ie different providers and their supported protocols. 
Noting that the three techniques \cite{anderson_tls_2019,fan_identify_2019,lastovicka_using_2020} that can achieve overall 80\% accuracy require significant adaptations from constructing flow-level telemetry to articulating attributes and developing classification models, whereas the technique in \cite{ren_app_2021} uses attributes extracted from flow metadata (\eg length) and only one TLS field ``\textit{TLS\_message\_type}'' which becomes unavailable/encrypted in QUIC, and thus has only 11.3\% accuracy for YouTube flows over QUIC and less than 60\% for other scenarios.

\section{Characterizing User Platforms in a Campus Network}
\label{sec:insights}
To showcase the usability of our system for residential broadband network operators, in \S\ref{sec:insights-prototype},
we report on the deployment of our user platform classification pipeline on a commodity server and use it to analyze traffic between our university campus network (mimicking a residential broadband network) and the Internet in real time. Then in \S\ref{sec:insights-insights}, we discuss insights obtained from our deployment over a four-month period, demonstrating how video streaming consumption patterns vary across user platforms and providers. 
Last, in \S\ref{sec:discussion}, we conclude discussing insights collected as well as general considerations from our deployment that might be of interest to ISPs' network operation teams interested in video streaming traffic analytics.

\subsection{Prototype and Campus Deployment}\label{sec:insights-prototype}
We implement our packet processing pipeline, shown in Fig.~\ref{fig:pipeline}, as a  virtual network function (VNF) system. 
DPDK \cite{dpdk} and NFF-Go \cite{a2022_aregmnffgo} packet processing frameworks are used for the packet \textit{Preprocessing} stage in Fig~\ref{fig:pipeline}, which parses input packet streams and handles the handshake and data packets of video flows. The handshake attribute generator and video telemetry modules are developed using Golang while the classifier banks use Python scikit-learn library \cite{scikit-learn}. The video session telemetry, along with user platform labels, is stored in a PostgreSQL database for analysis. 

The system runs on a commodity blade server configured with an 8-core Intel Xeon E5-2620 CPU and 64 GB DDR4 RAM. The server receives a copy of the traffic from our university network border router that is connected to the Internet. The traffic is delivered to two 10 Gbps network interfaces on our server for inbound and outbound traffic, respectively. We have obtained ethical clearance from our university ethics board as detailed in Appendix~\ref{sec:Appendix-Ethics}.

As an additional sanity check of the classification methodology, we played over 1000 video sessions using all available user platforms in our university lab, which were captured by our system. The classification accuracy of those ground-truth video sessions is similar to our open-set evaluation. Almost all misclassified instances (which are less than 4\% of streams) were with relatively low confidence, \ie < 50\%.

\begin{figure}[!t]
	\vspace{3mm}
	\includegraphics[width=\linewidth]{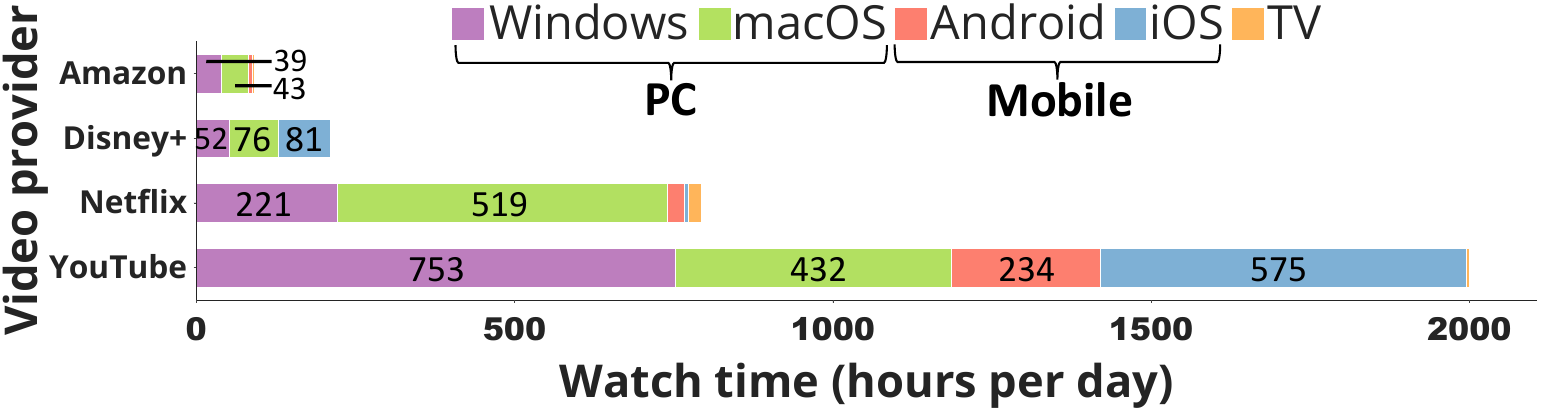}
	\vspace{-3mm}
	\caption{Video watch time for the four video content providers across device types.}
	\label{fig:watchtime_all}
\end{figure}

\subsection{Insights from the Deployment}\label{sec:insights-insights}
Our university campus network serves staff, students, visitors, and over 10 residential dormitories. Over the four months from July 7th 2023, 00:00 to November 9th 2023, 23:59, our system collected telemetry statistics, \ie duration, volume and throughput, of every video flow from the four providers, and tagged each of them by their user platforms.
Over 100 million streaming video flows were collected with a total of 400k hours of watch time. For reliability of our insights, we exclude about 20\% of the sessions with low classification confidence that may be due to unknown types of user platforms not in our training dataset.
We now analyze how different user platforms impact usage patterns from the point of view of watch time (\S\ref{sec:insights-aggregated}), bandwidth demand (\S\ref{sec:insights-bandwidth}), and temporal usage patterns (\S\ref{sec:insights-temporal}).

\subsubsection{Video watch time across user platforms}\label{sec:insights-aggregated}
Taking our four-month deployment as an example, we start by examining how the level of engagement, measured in terms of total watch time, varies by device type across the four video content providers. This is shown in Fig.~\ref{fig:watchtime_all}.
Not surprisingly, across our entire campus demographic, YouTube, where content is mostly free, dominates engagement with an average daily total watch time of 2000 hours. It is followed by subscription-based providers such as Netflix, Disney+, and Amazon Prime Video.
Furthermore, the majority of subscription-based videos are watched on PCs (Windows/Mac) rather than on mobile devices. In contrast, up to 40\% of YouTube engagement occurs on mobile devices (iOS/Android). These insights enable ISPs to prioritize the troubleshooting of issues, such as they can expect more support calls relating to PCs than mobiles for Netflix, and the converse for YouTube.

\begin{figure}[!t]
	\mbox{
		\hspace{-2mm}
		\subfigure[YouTube.]{
			{\includegraphics[width=0.49\linewidth]{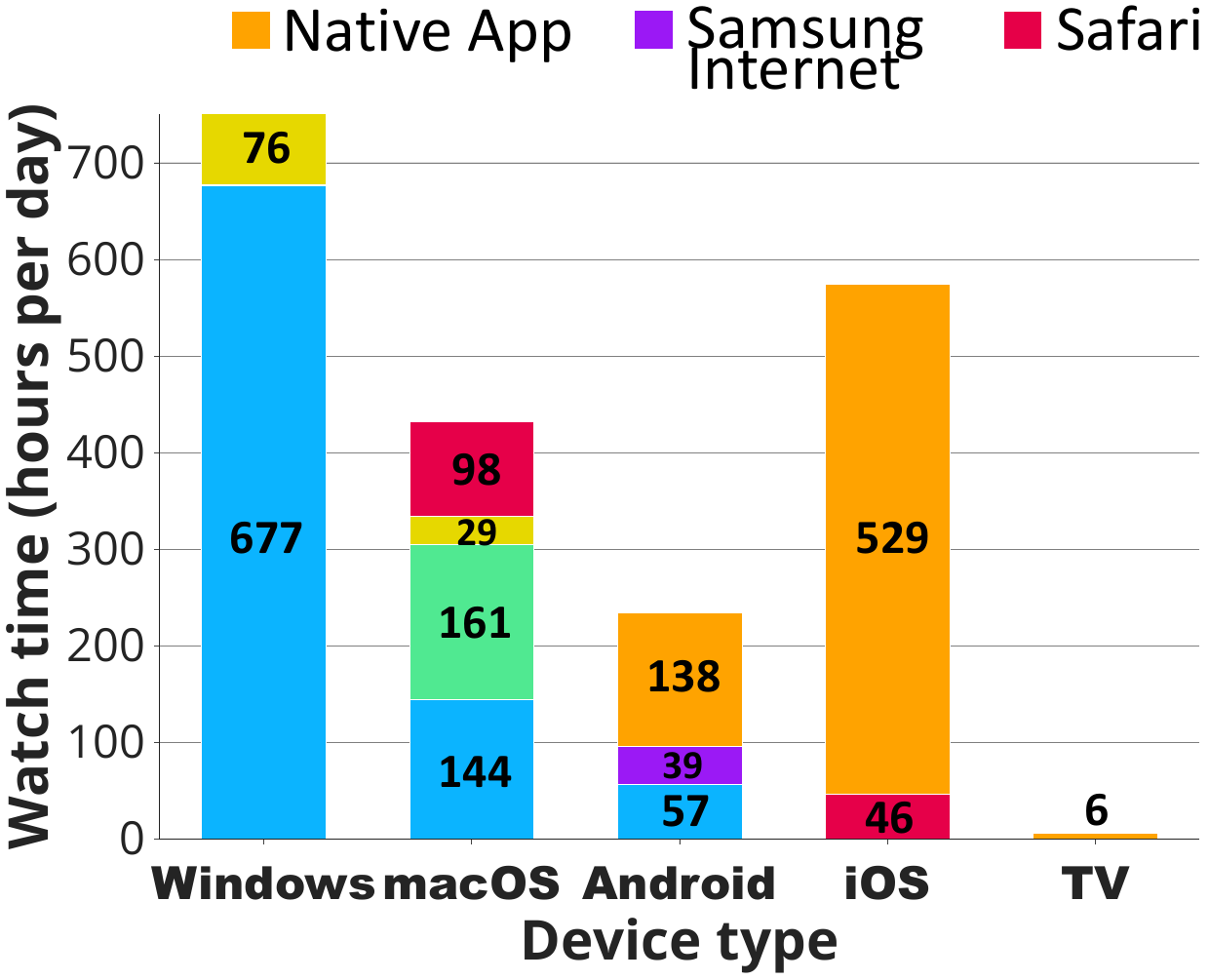}}\quad
			\label{fig:watchtime_youtube}
		}
		\hspace{-5mm}
		\subfigure[Netflix.]{
			{\includegraphics[width=0.49\linewidth]{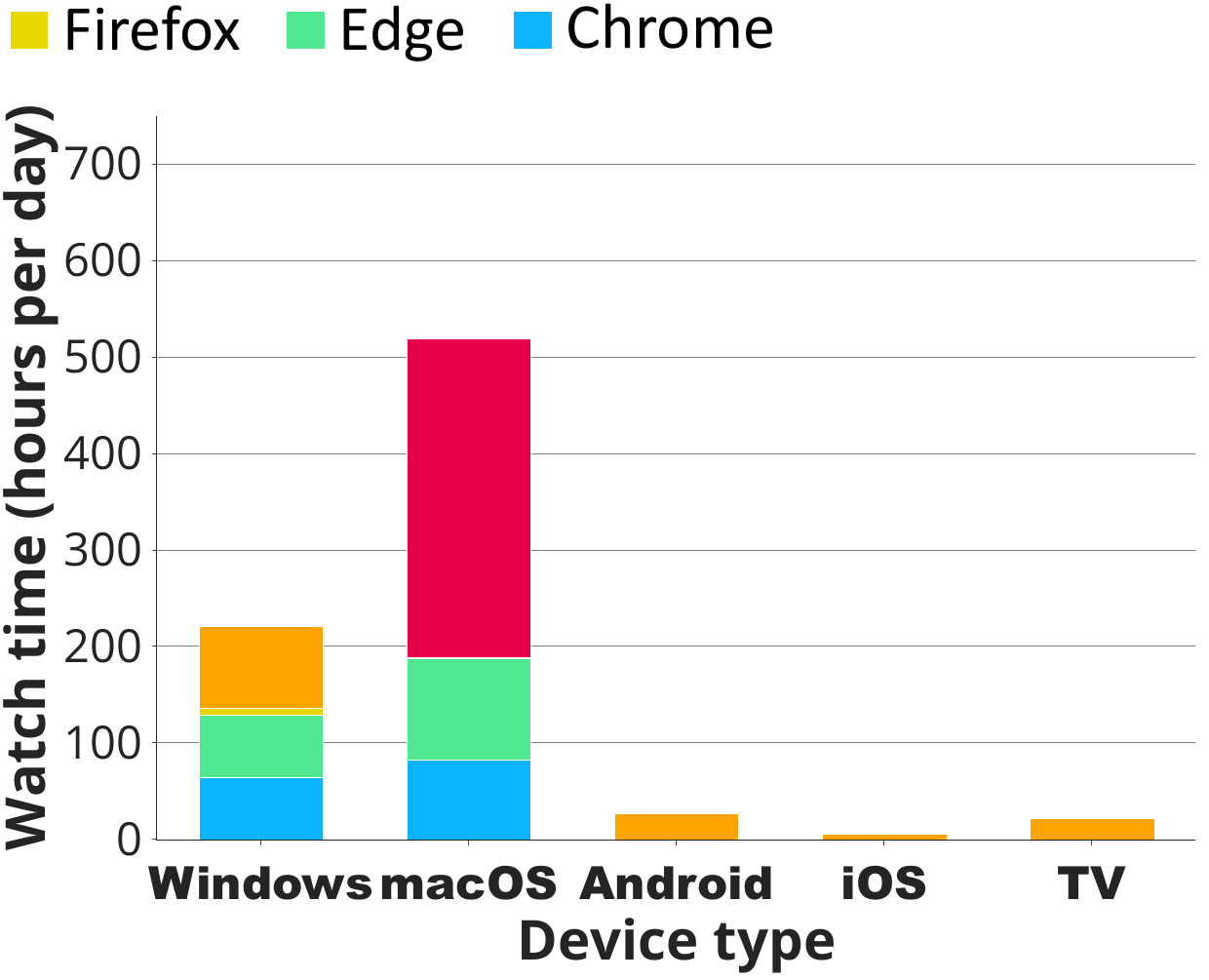}}\quad
			\label{fig:watchtime_netflix}
		}
	}
	\mbox{
		\hspace{-2mm}
		\subfigure[Disney+.]{
			{\includegraphics[width=0.49\linewidth]{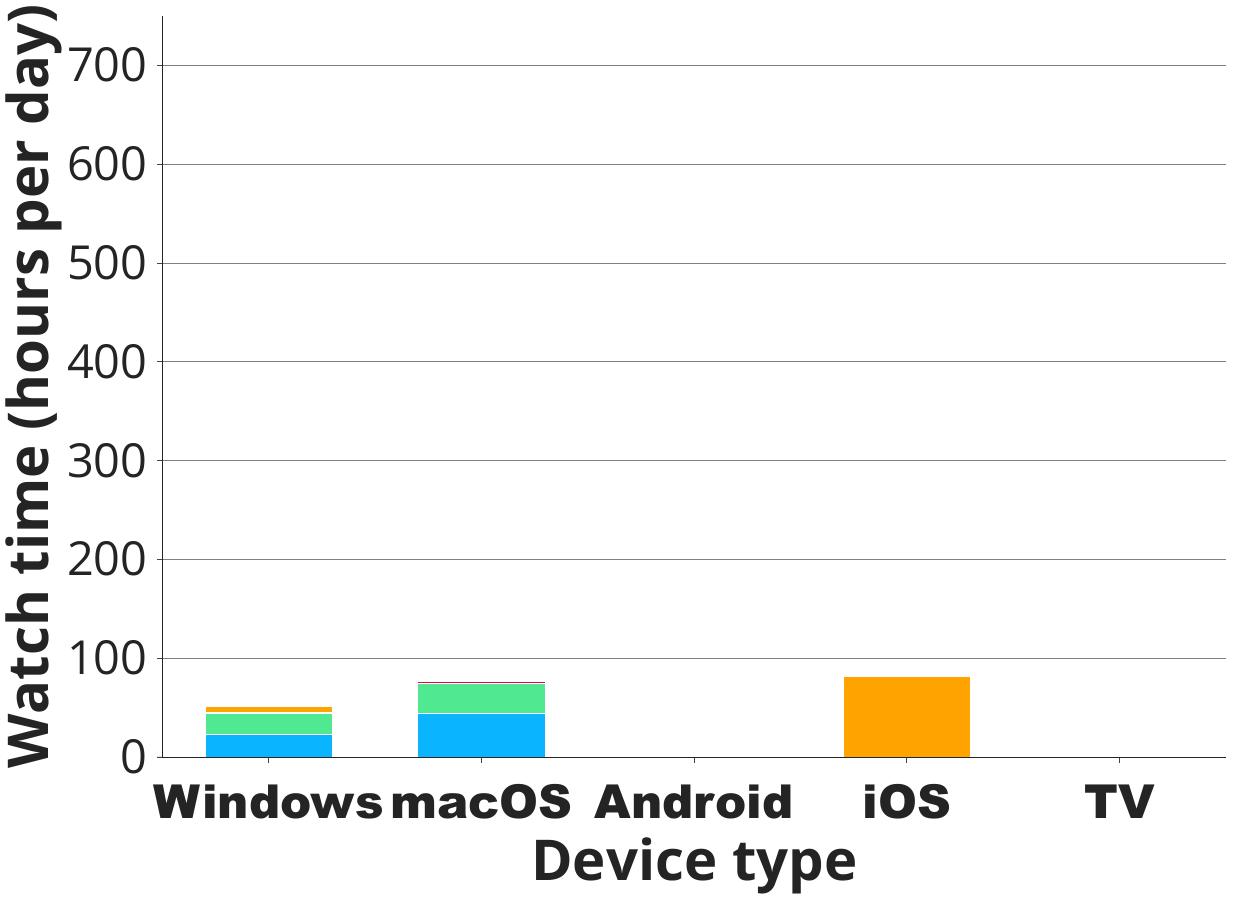}}\quad
			\label{fig:watchtime_disney}
		}
		\hspace{-5mm}
		\subfigure[Amazon Prime Video.]{
			{\includegraphics[width=0.49\linewidth]{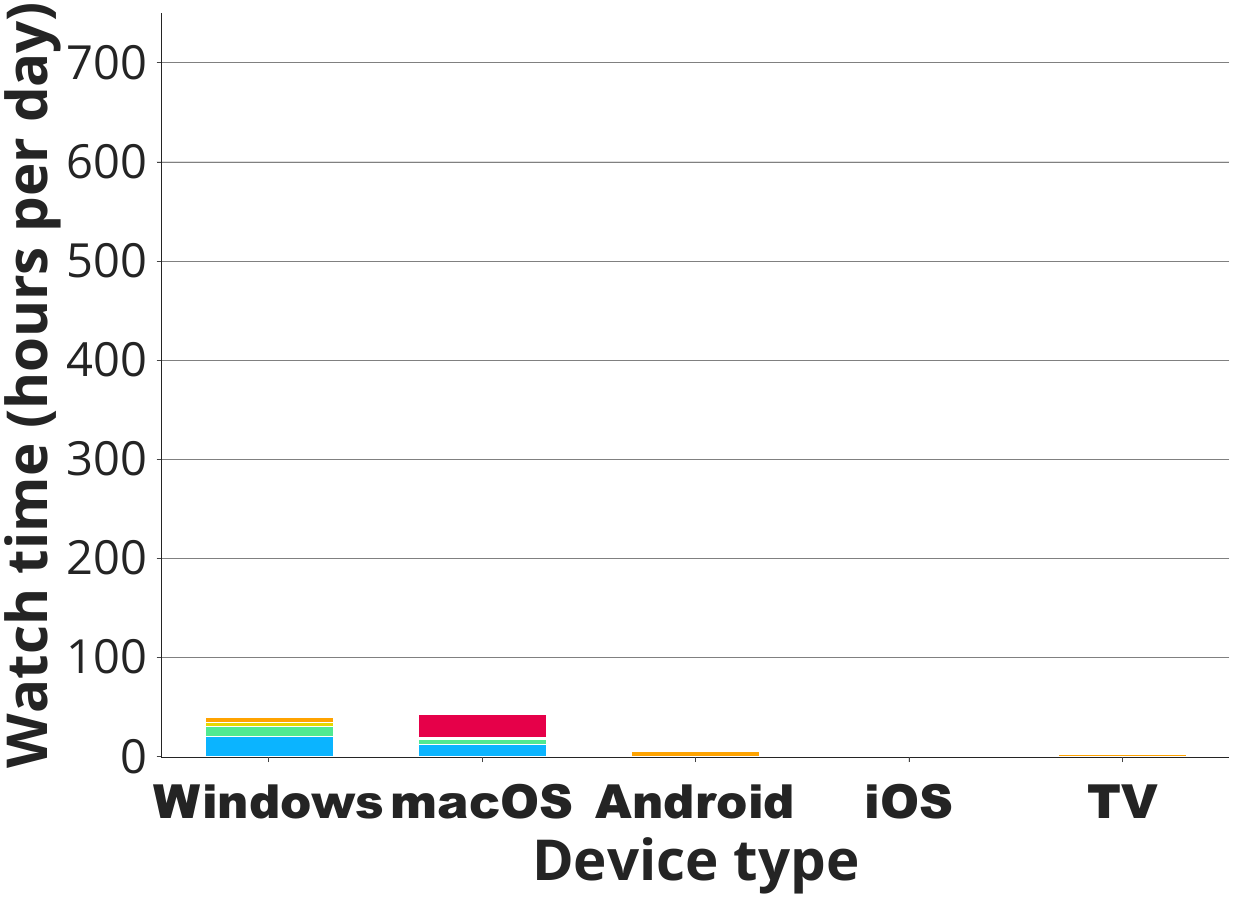}}\quad
			\label{fig:watchtime_amazon}
		}       
	} 
	\vspace{-3mm}
	\caption{Video watch time for the four video content providers across software agents on each device type.}
	\label{fig:insights_overview}
\end{figure}

Fig.~\ref{fig:insights_overview} shows the breakdown of software agents per video provider.
Chrome browser on Windows PCs is the most popular software agent used to watch YouTube, clocking up 677 hours,  as shown in Fig.~\ref{fig:watchtime_youtube}. In addition, amongst mobiles, iOS is preferred with over 90\% of watch time on its YouTube native app. The other devices use a relatively diverse set of software agents, as shown in the figure.  

The watch time profiles for Netflix, Disney+ and Amazon are provided in Fig.~\ref{fig:watchtime_netflix}, \ref{fig:watchtime_disney} and \ref{fig:watchtime_amazon}, respectively. While Safari on Mac PCs is popular for viewing Netflix and Amazon, the native Disney+ app on iOS dominates engagement of mobile users by over 90\%. 

\begin{figure}[!t]
	\vspace{2mm}
	\includegraphics[width=\linewidth]{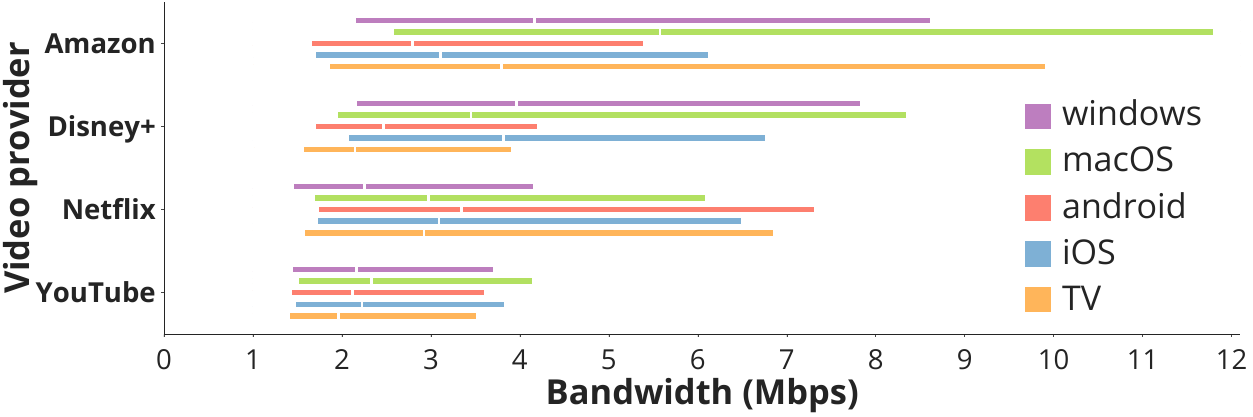}
   \vspace{-3mm}
	\caption{Bandwidth demand for the four video providers across device types.}
	\label{fig:bandwidth_overview}
\end{figure}

\begin{figure}[!t]
	\mbox{
		\hspace{-3mm}
		\subfigure[Amazon Prime Video.]{
			{\includegraphics[width=0.5\linewidth]{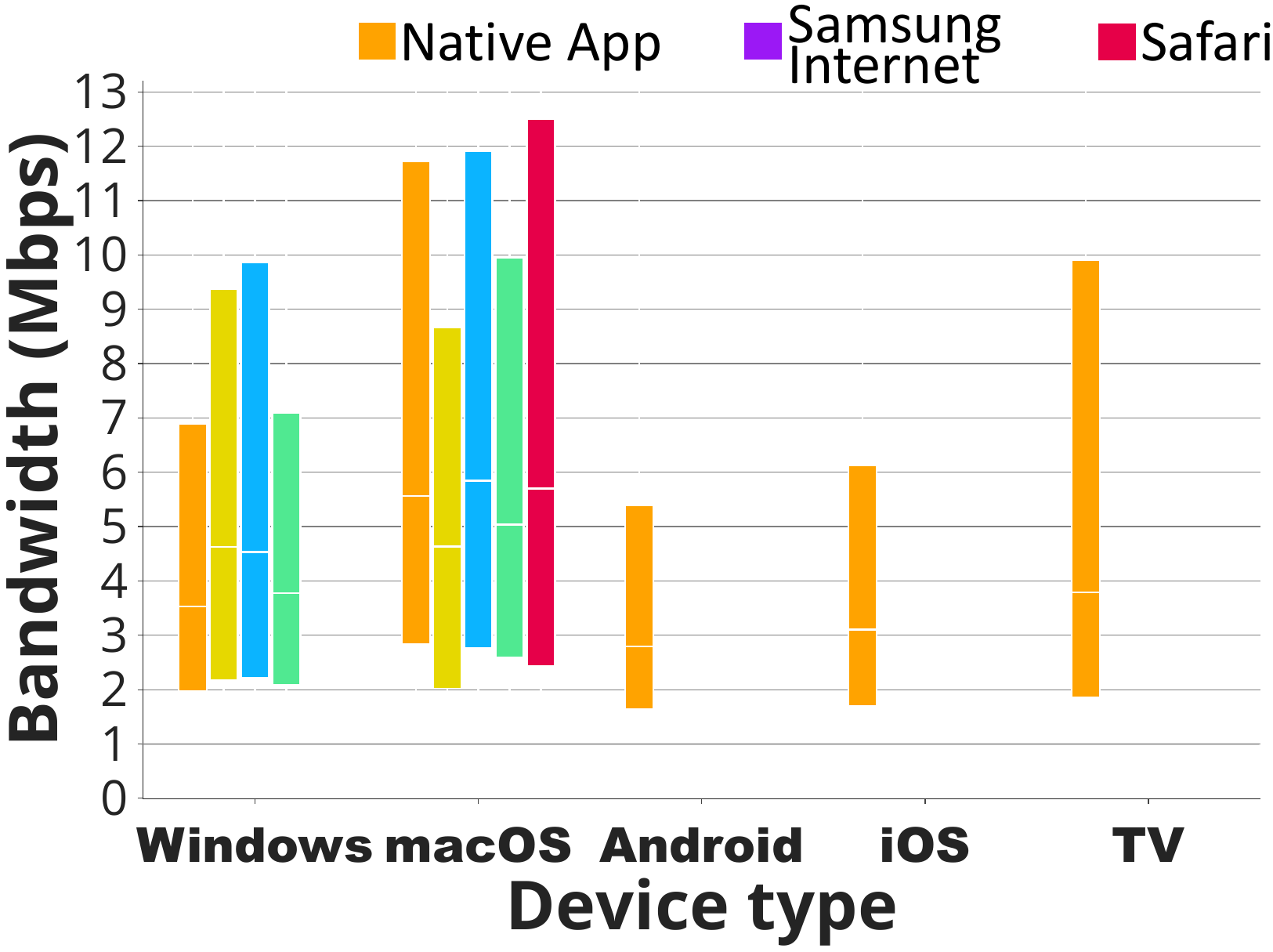}}\quad
			\label{fig:bandwidth_amazon}
		}
		\hspace{-5mm}
		\subfigure[Disney+.]{
			{\includegraphics[width=0.5\linewidth]{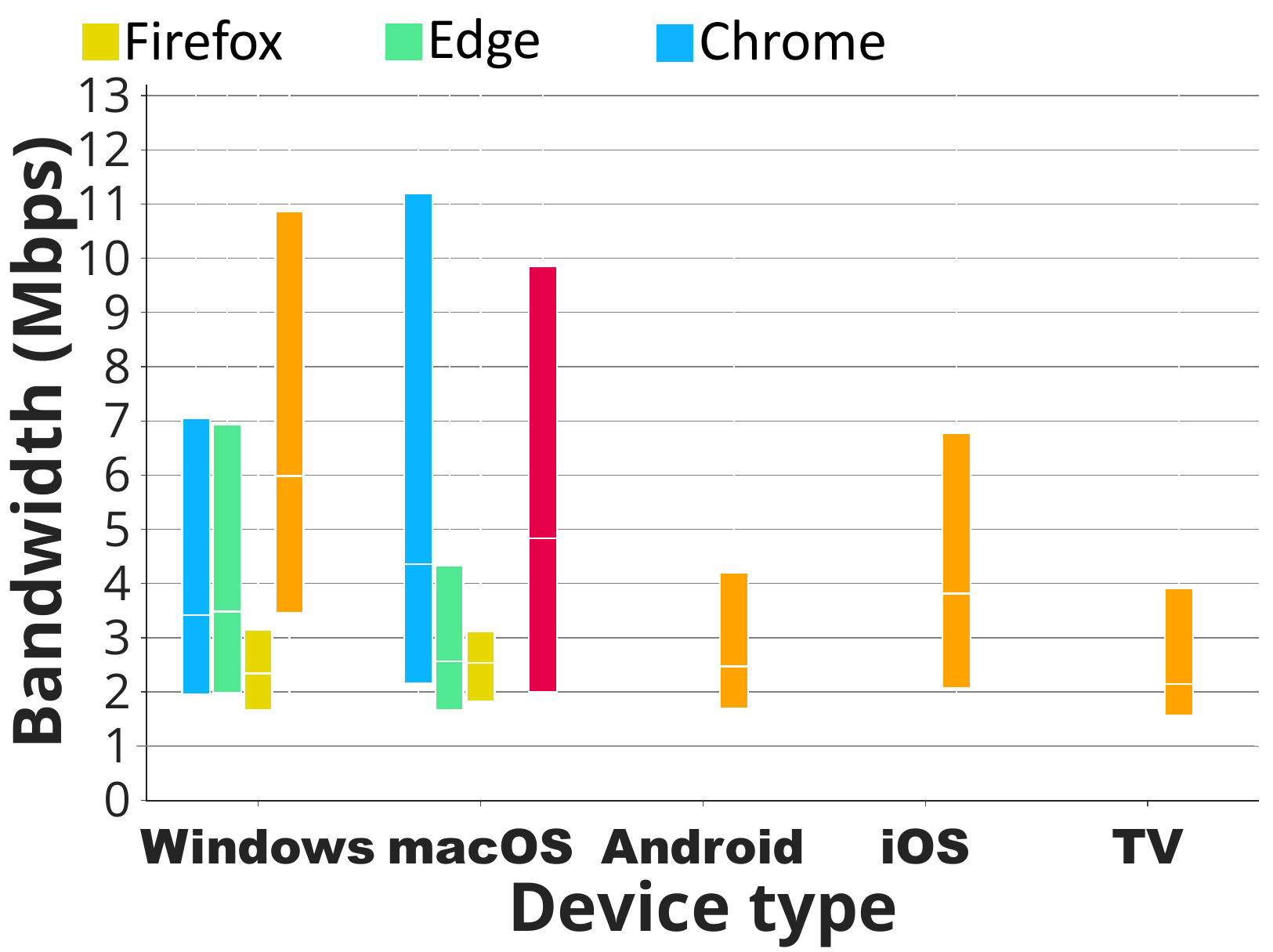}}\quad
			\label{fig:bandwidth_disney}
		}
	}
	\mbox{
		\hspace{-3mm}
		\subfigure[Netflix.]{
			{\includegraphics[width=0.5\linewidth]{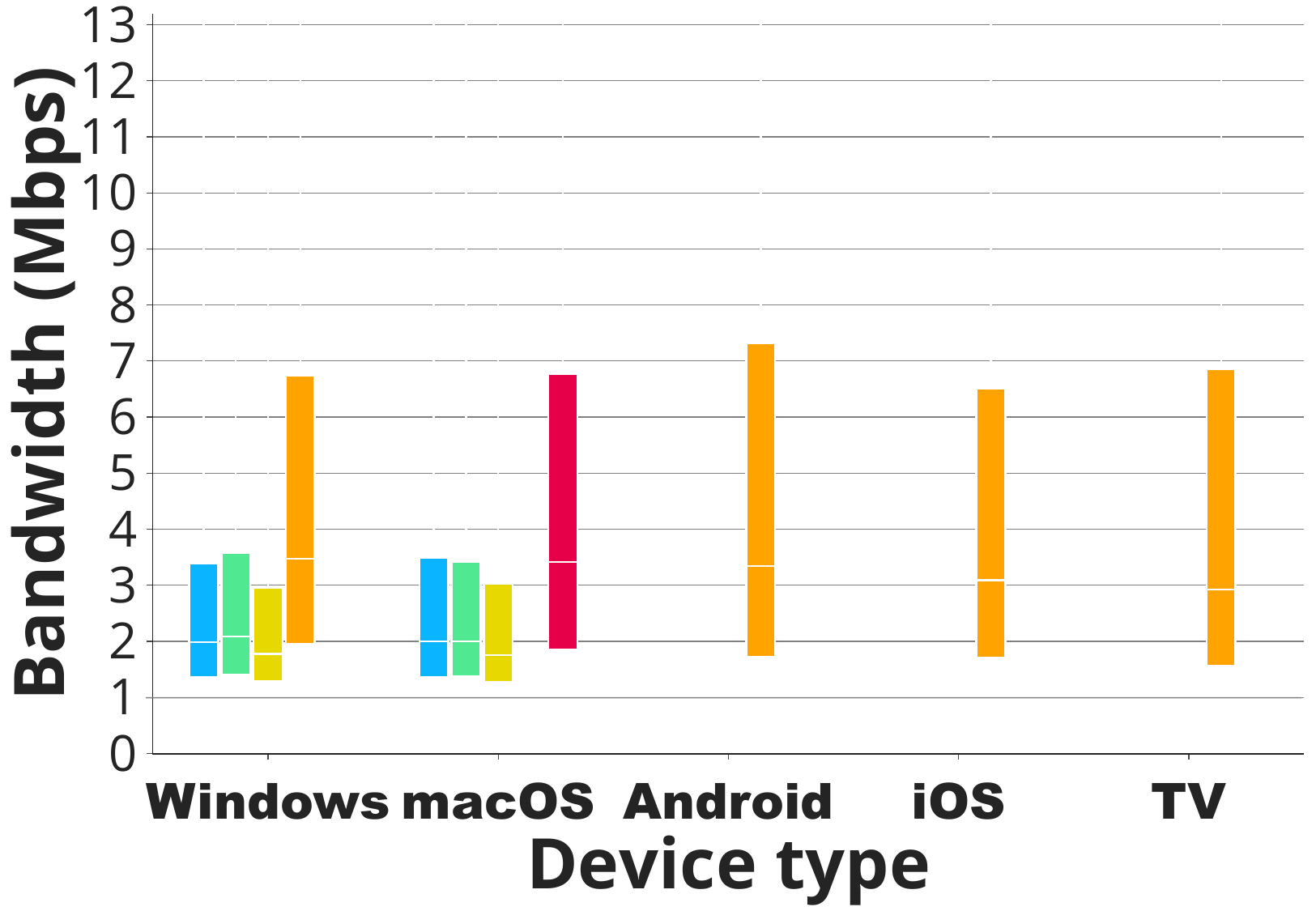}}\quad
			\label{fig:bandwidth_netflix}
		}
		\hspace{-5mm}
		\subfigure[YouTube.]{
			{\includegraphics[width=0.5\linewidth]{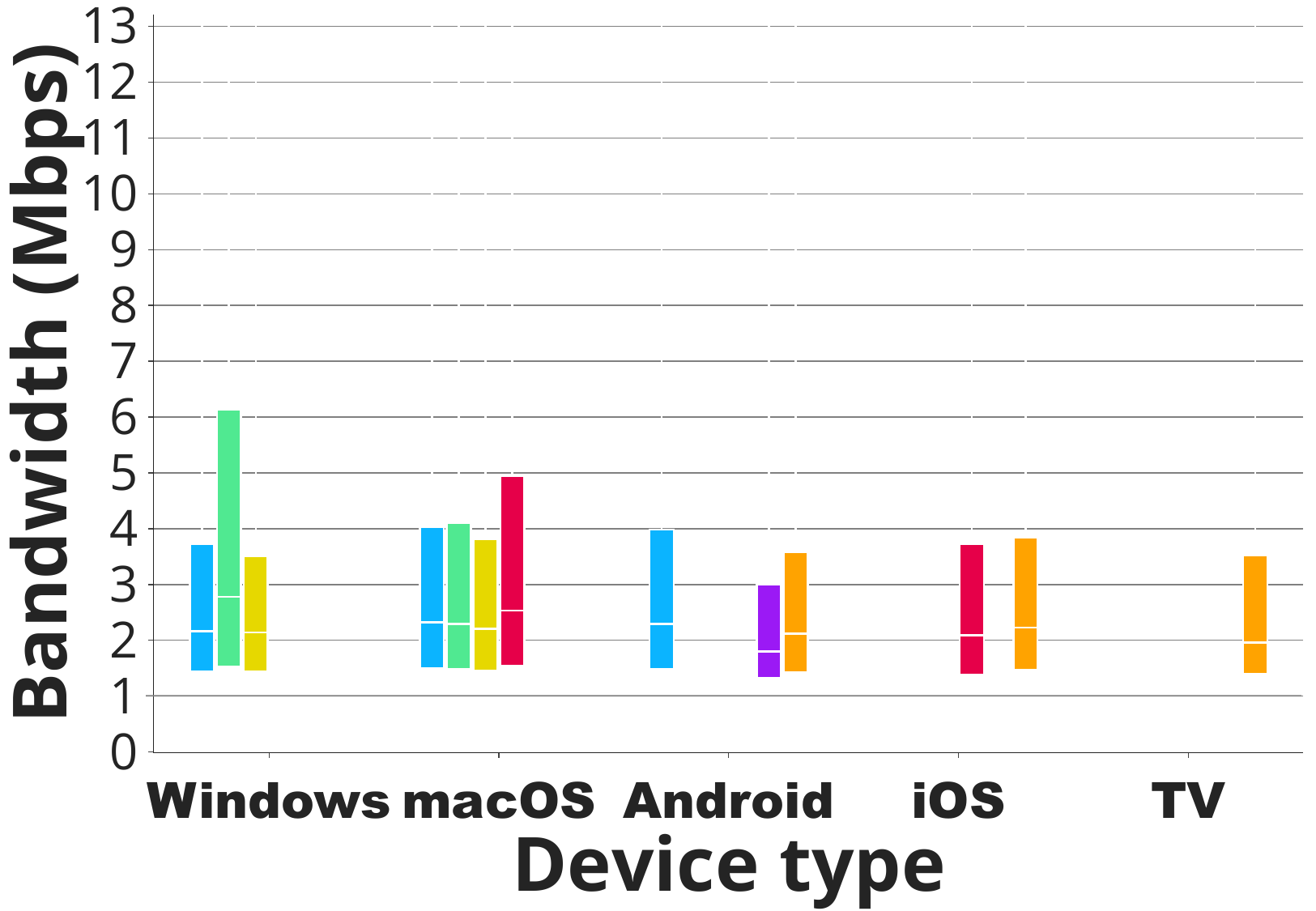}}\quad
			\label{fig:bandwidth_youtube}
		}
	}
	        \vspace{-3mm}
	\caption{Bandwidth demand for the four video providers across software agents on each device type.}
	\label{fig:bandwidth_deployment}
\end{figure}

\subsubsection{Bandwidth demand}\label{sec:insights-bandwidth}
As video streaming are bandwidth demanding, in Fig.~\ref{fig:bandwidth_overview}, we show the distribution of bandwidth consumption, as box plots indicating the median and the quartiles to either side of the median, for the four video streaming providers across different device types. It is apparent that the bandwidth demand imposed by subscription-based videos is higher than that of YouTube as the interquartile range for these providers is 3 to 9 Mbps higher. Notably, videos streamed from Amazon Prime Video to Mac PCs demand the highest median bandwidth (of 5.7 Mbps), which is 50\% higher compared to smart TVs. 

The impact of software agents on bandwidth for the different video providers is plotted in Figures~\ref{fig:bandwidth_amazon}-\subref{fig:bandwidth_youtube}. Android mobiles, iOS mobiles and TVs only support native Amazon Prime Video apps and as shown in Figures~\ref{fig:bandwidth_amazon}, consume less bandwidth (median $<$ 3 Mbps) than their PC counterparts. All browsers on Windows/Mac PCs exhibit higher median bandwidth and interquartile range spread for Amazon compared to native mobile apps. In addition, Mac PCs generally require higher median throughput than Windows PCs.
Interestingly, Netflix streamed to PCs on browsers (excluding Safari) consumes lower median bandwidth ($<$ 2 Mbps), which might suggest lower resolution supported via browsers than the native app. 

\subsubsection{Temporal usage patterns}\label{sec:insights-temporal}
\begin{figure}[!t]
	\mbox{
		\hspace{-3mm}
		\subfigure[Amazon Prime Video.]{
			{\includegraphics[width=0.5\linewidth]{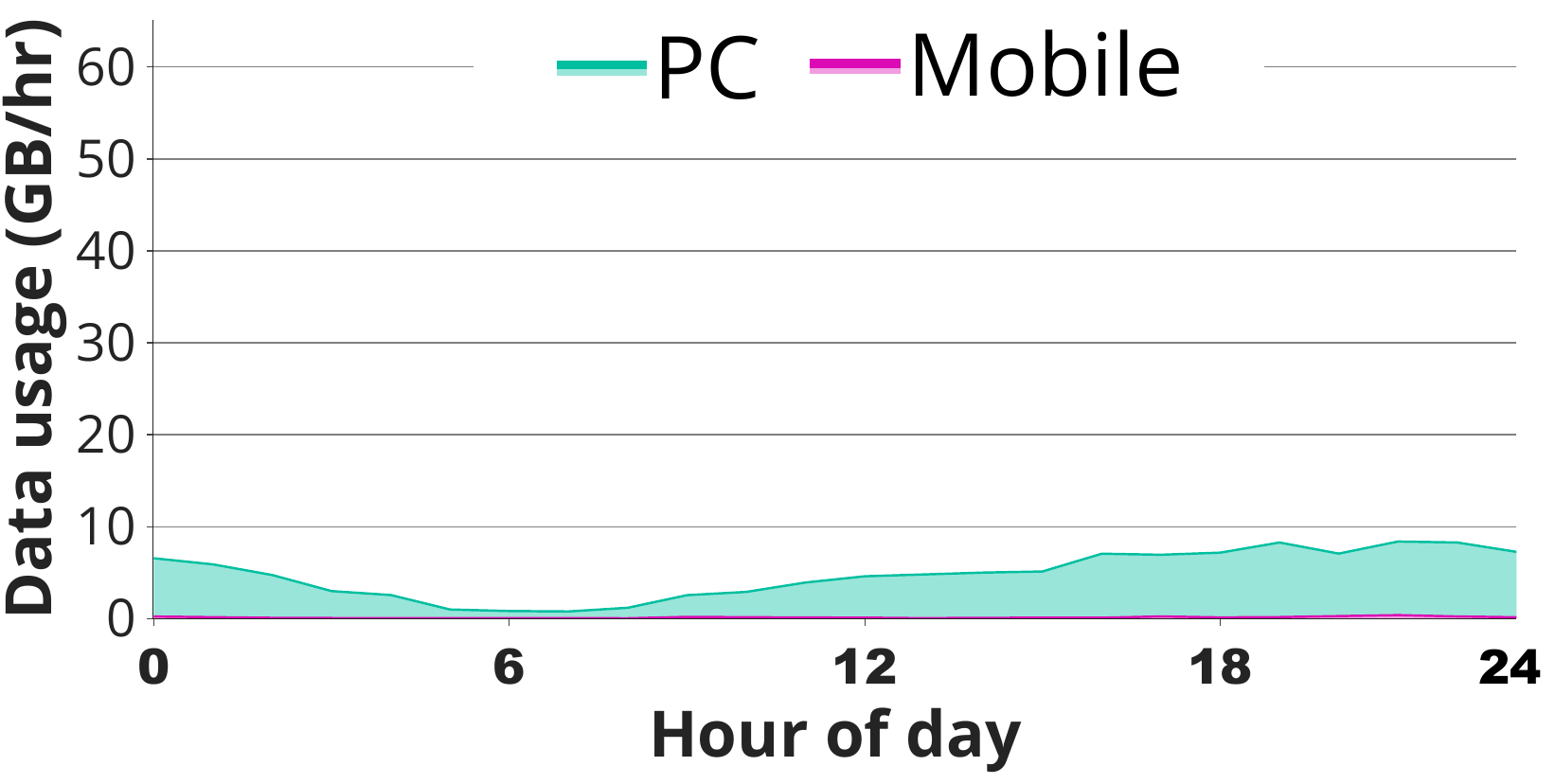}}\quad
			\label{fig:temporal_amazon}
		}
		\hspace{-5mm}
		\subfigure[Disney+.]{
			{\includegraphics[width=0.5\linewidth]{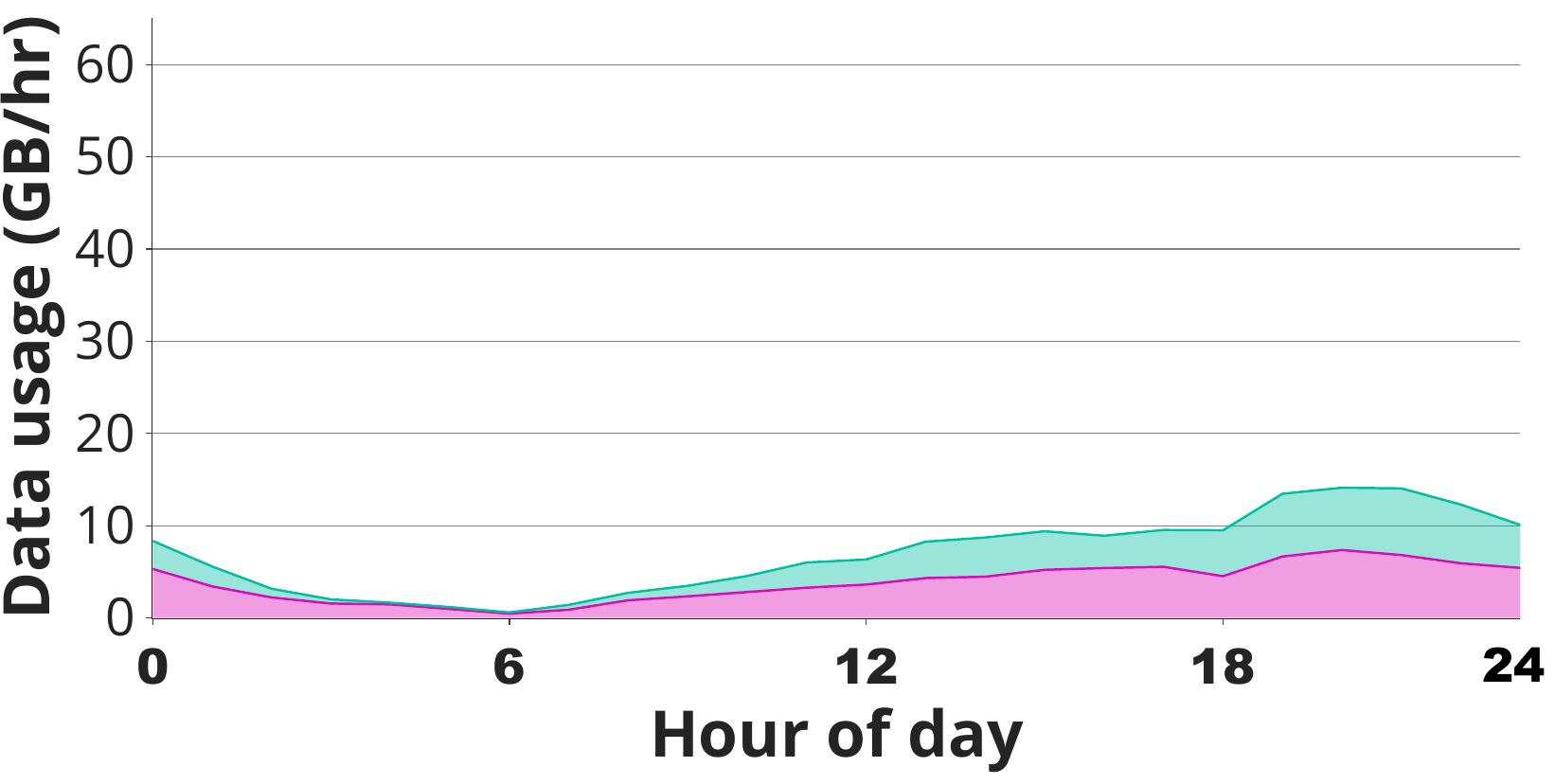}}\quad
			\label{fig:temporal_disney}
		}
	}
	\mbox{  
		    \hspace{-3mm}
		\subfigure[Netflix.]{
			{\includegraphics[width=0.5\linewidth]{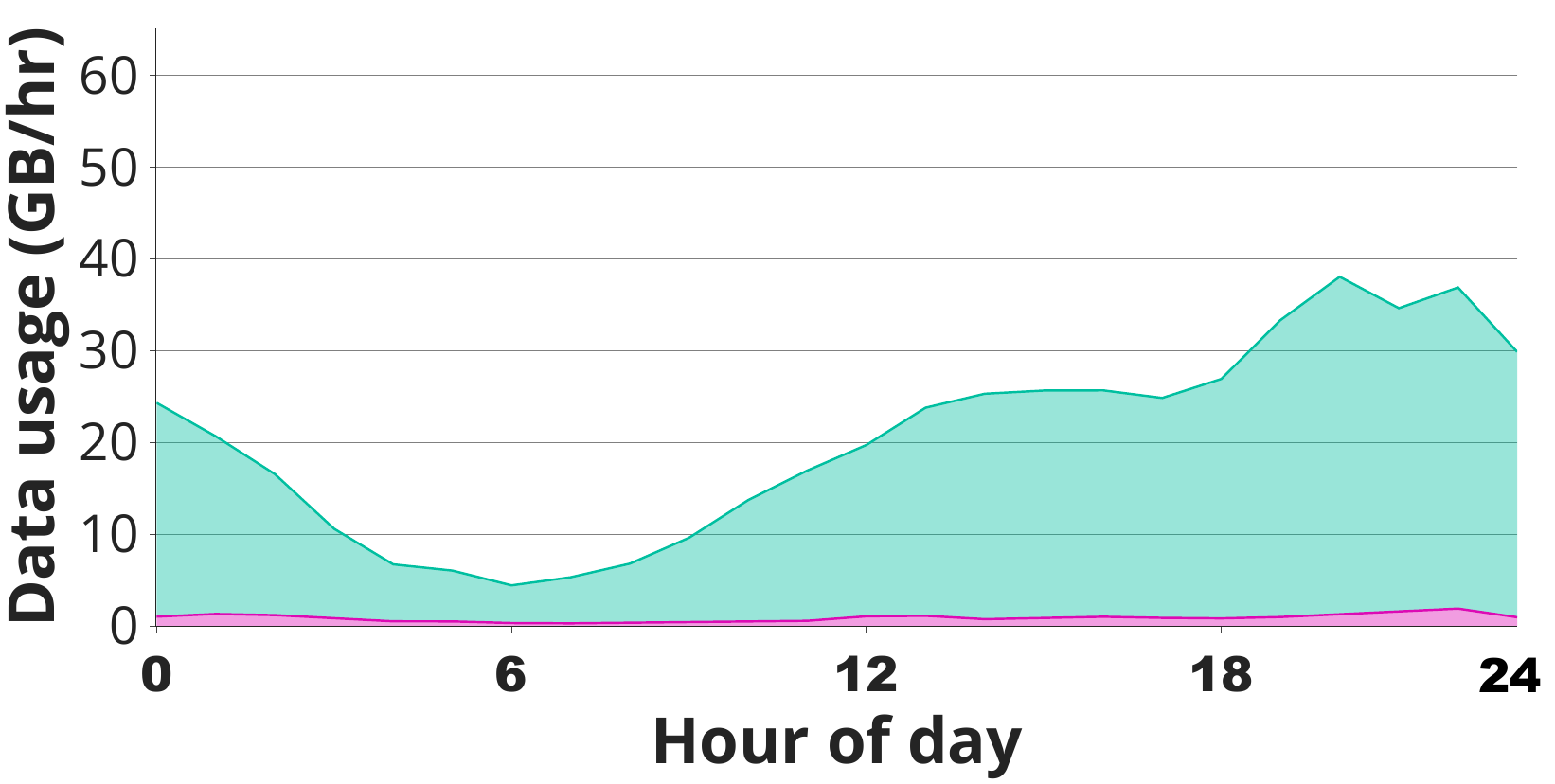}}\quad
			\label{fig:temporal_netflix}
		}
		\hspace{-5mm}
		\subfigure[YouTube.]{
			{\includegraphics[width=0.5\linewidth]{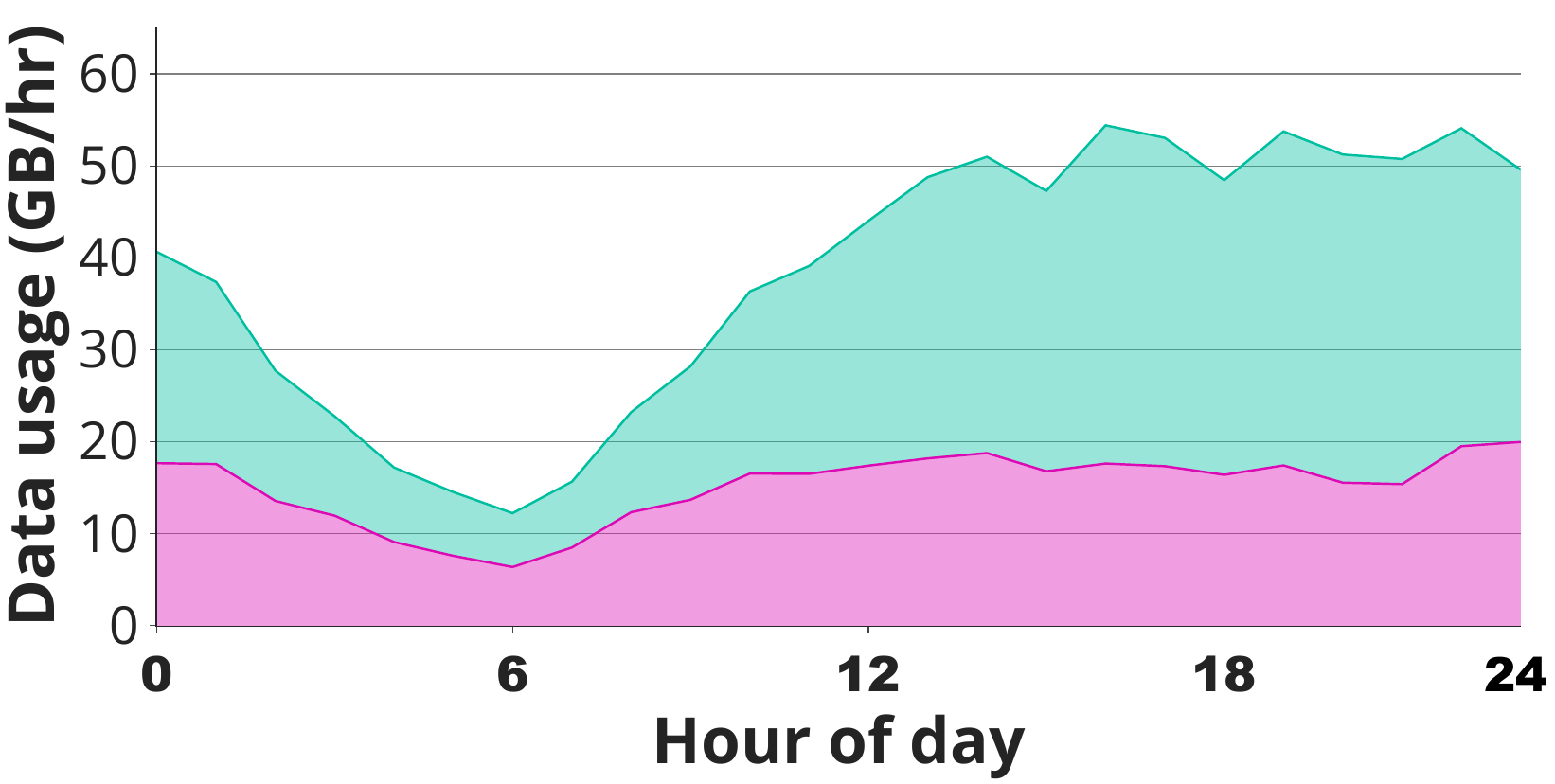}}\quad
			\label{fig:temporal_youtube}
		}
	}
	      \vspace{-3mm}
	\caption{Temporal data usage patterns for the four content providers across PCs and mobile devices.}
	\label{fig:temporal_deployment}
\end{figure}

As traffic demand changes over time, in Fig.~\ref{fig:temporal_deployment}, we depict the median traffic volume during each hour of the day consumed by Amazon, Disney+, Netflix and YouTube videos on PCs and mobile devices. Overall, Amazon and Disney+ exhibit fairly similar daily usage patterns with a 4-hour peak period from about 7 pm to 11 pm. Also, mobile usage for Amazon is low compared to Disney+.

Comparing YouTube and Netflix, we see that the former has a long and sustained peak window from about 4 pm to midnight while the latter has a shorter peak, \ie between 8 pm to 10 pm. In terms of mobile usage, YouTube dominates with relatively steady hourly peak usage in the range of 17 to 20 GB from about 4 pm to midnight.

\subsection{Discussion on Insights and Considerations}\label{sec:discussion}
For network operation teams interested in video streaming, visibility into engagement patterns across user platforms can equip them with knowledge of their customer segments and preferences, as well as the ability to identify customer issues pertinent to popular content providers and/or certain firmware/software. 
Second, bandwidth capacity planning is a key ongoing activity for ISPs as they strive to meet the ever-increasing demands for service quality assurance from their customers. Given the significant load imposed by video streaming, fine-grained insights into the bandwidth demand of popular streaming service users, broken down by device types and software agents, may help ISPs improve the fidelity of their bandwidth forecasting models. 
In addition, temporal usage patterns of video streaming services can also provide valuable information to ISPs. For example, by knowing when peak usage occurs and for what kind of content, ISPs can proactively allocate adequate bandwidth to ensure high levels of service assurance. Conversely, during off-peak periods, bandwidth can be allocated differently to save costs. Alternatively, traffic management policies can be implemented that prioritize video streaming during peak hours and other types of traffic during off-peak hours. 

We acknowledge that our study focused particularly on four popular content providers and thirty user platforms to demonstrate the efficacy of our method. While it is a necessary engineering extension to continuously collect comprehensive datasets and retrain models, the methodology itself is readily extensible to incorporate other streaming services, emerging user platforms, and changes in behaviors over time.
Therefore, in practical deployment for a broadband network, two key \textbf{deployment considerations} arise. 

First, for deployments in different broadband networks, the operators can have various video streaming services of interest. For example, a deployment in Asia-Pacific region may require the system to be able to measure streaming video sessions from regionally popular providers such as Bilibili and Hotstar. Therefore, the deployment team would have to collect ground-truth data of video streaming flows for each video provider to be included. Similarly, new popular user platforms for video streaming may emerge. For example, broadband ISPs are interested in understanding the streaming videos watched on Apple Vision Pro after it was released in February 2024, which requires the collection of ground-truth data to augment the classifiers.

Second, although our four-month deployment did not exhibit any observable drop in prediction confidence from our classifiers over time, it is acknowledged that the overall prediction accuracy and confidence will decline over a longer deployment period due to evolving traffic characteristics of video streaming services and (firmware/software) updates of user platforms, which is known as ``concept drift''. Therefore, in practice, the deployment team will have to periodically retrain the under-performing classifiers with updated ground-truth training data to adapt to these changes. While developing a continuous re-training process is not within the scope of this paper, we acknowledge that there are established techniques to detect and mitigate concept drifts \cite{sheluhin_concept_2021,malekghaini_data_2022,malekghaini_deep_2023}.

\section{Conclusion}\label{sec:conclusion}
Our work discussed in this paper provides network operators with fine-grained visibility into user platforms of streaming video flows over both TCP and QUIC.  
We first systematically understand the network communication anatomy of streaming video sessions and categorize handshake fields of video flows that can vary across OSes, browsers and provider-native applications. Following our observation on the variations of handshake fields across user platforms, we developed and evaluated a pipeline that processes network packet streams to classify user platforms of streaming video flows using well-trained machine learning models on formalized attributes from handshake fields of video flows, achieving over 96\% accuracy.
We then prototype our system on a commodity server and deploy it in a large university campus network that mimics a residential broadband network for a four-month period. The usage patterns from over 100 million video flows across various user platforms accessing four major content providers, namely YouTube, Netflix, Amazon Prime Video and Disney+ are discussed. Our method provides ISPs with valuable insights to better understand their customer segments, provision bandwidth, and troubleshoot video streaming issues pertinent to device firmware, OS, or software for customer experience and satisfaction.

\section*{Acknowledgments}
We thank our shepherd Alessandro Finamore and the five anonymous reviewers for their comprehensive and insightful feedback. This work is supported by Australian Government’s National Industry PhD Program award reference number 35063 and Cooperative Research Centres Projects (CRC-P) Grant CRCPXIV000099.

\bibliographystyle{ACM-Reference-Format}
\balance
\bibliography{reference}

\appendix

\section{Ethics}\label{sec:Appendix-Ethics}
 We have obtained ethical clearance from our university ethics board (UNSW Human Research Ethics Advisory Panel approval number HC211007) that allows us to analyze campus traffic for streaming video flows without being privy to user identities such as ID numbers and names.
In our campus deployment, insights into video streaming user platforms were reported in an aggregated manner, preserving anonymity rather than identifying specific users. In our analysis, no attempt was made to associate video streaming sessions with personal identities.

\section{Handshake Field Values of Video Flows}\label{sec:Appendix-Handshake}
In this section, we show the distribution of handshake field values across all user platforms for all four considered streaming services (\ie YouTube, Netflix, Disney and Amazon).

\begin{figure}[H]
	\subfigure[Handshake field values of \textbf{\color{alizarin}YouTube} flows over QUIC across 12 user platforms.]{
		\includegraphics[width=0.48\textwidth]{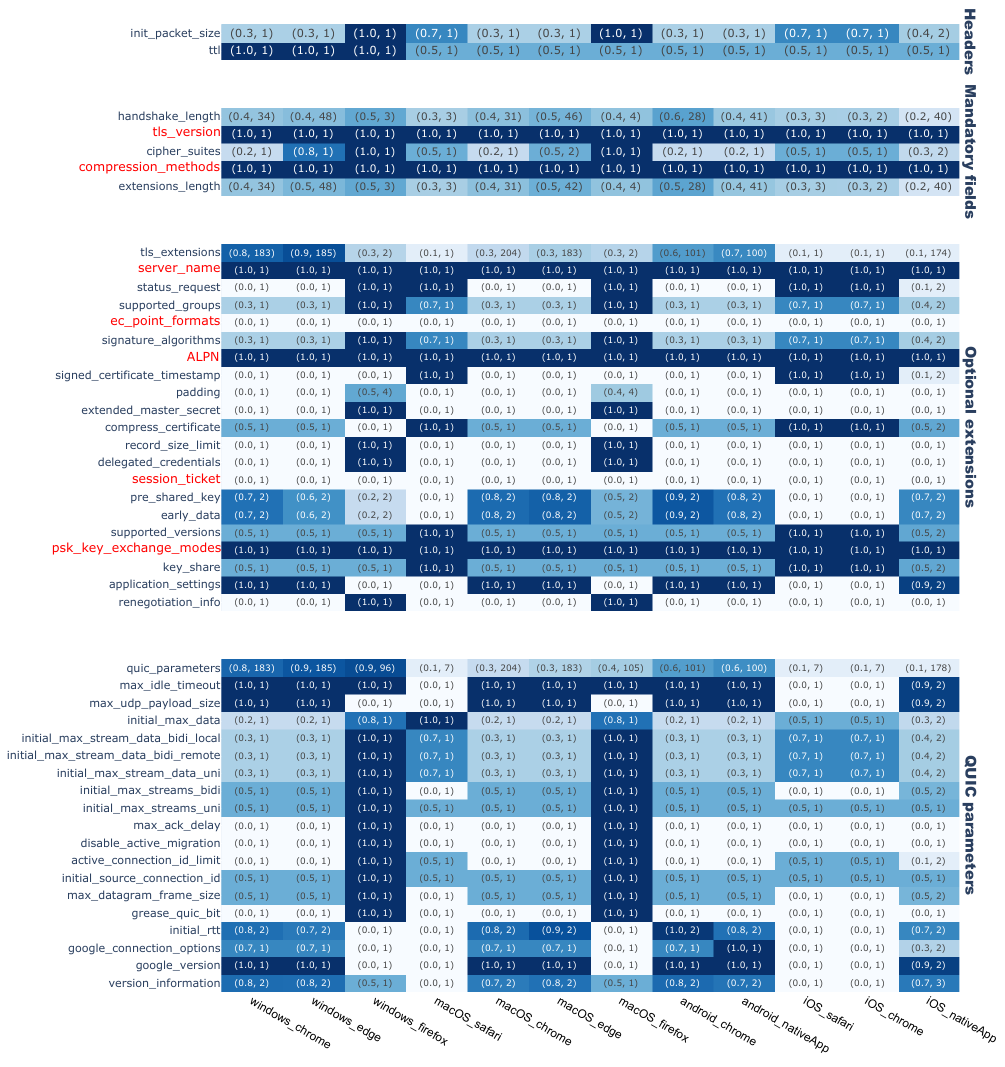}
		\label{fig:heatmap_quic}}
	\subfigure[Handshake field values of \textbf{\color{alizarin}YouTube} flows over TCP across 14 user platforms.]{
		\includegraphics[width=0.48\textwidth]{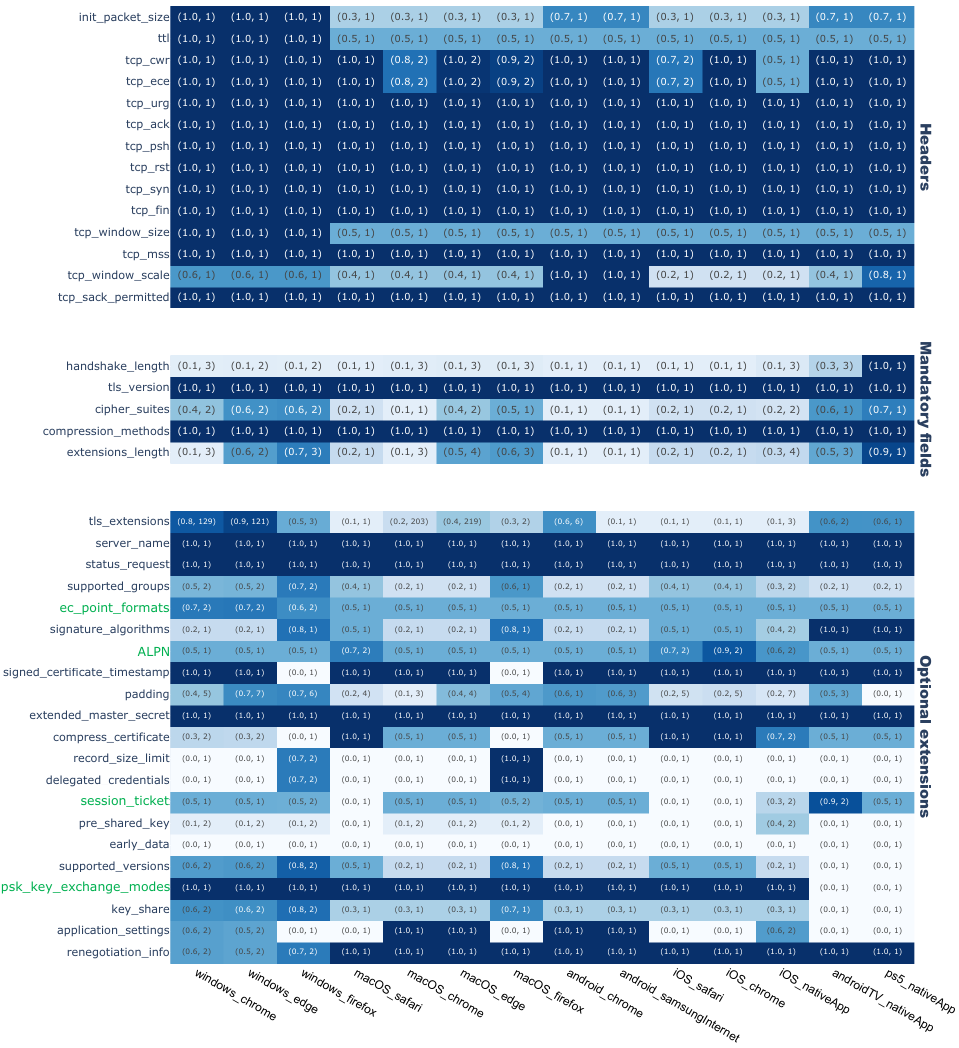}
		\label{fig:heatmap_tcp}}
	\caption{Median (normalized) and number of unique values shown as (x, y) taken by fields in handshake messages for {\color{alizarin}YouTube} flows.}
\end{figure}

In Fig.~\ref{fig:heatmap_quic}, we show a heatmap where each cell, corresponding to a user platform (indicated at the bottom of the figure), is a two tuple (x, y); x is the median value of the field, shown via labels on the left-hand side of the figure, and y is the number of distinct values that field takes as seen in our dataset. The heatmap is based on YouTube over QUIC flows across 12 user platforms. We consider this example for illustrative purposes. 

The median field values are normalized between 0 and 1. The x-axis shows 12 combinations of user platforms and the y-axis depicts the important fields obtained from different categories, as explained above. 

As shown in Fig.~\ref{fig:field_distribution_quic}, there are 7 fields whose median values are all the same (consistently either 0 or 1) across user platforms. These are highlighted as red labels in the figure and include \textit{tls\_version}, \textit{compression\_methods}, \textit{server\_name}, \textit{ec\_point\_formats}, \textit{ALPN}, \textit{session\_ticket} and \textit{psk\_key\_exchange\_modes}. This means they are not useful in differentiating user platforms for YouTube video flows over QUIC. However, as shown in green in Fig.~\ref{fig:heatmap_tcp}, 4 of these fields, \ie \textit{ec\_point\_formats}, \textit{ALPN}, \textit{session\_ticket} and \textit{psk\_key\_exchange\_modes} take different values across user platforms for YouTube video flows over TCP, meaning they can serve as useful indicators in identifying specific user platforms.

\begin{figure}[H]
	\subfigure[Handshake field value distribution of \textbf{\color{blue-violet}Netflix} flows.]{
		\includegraphics[width=0.47\textwidth]{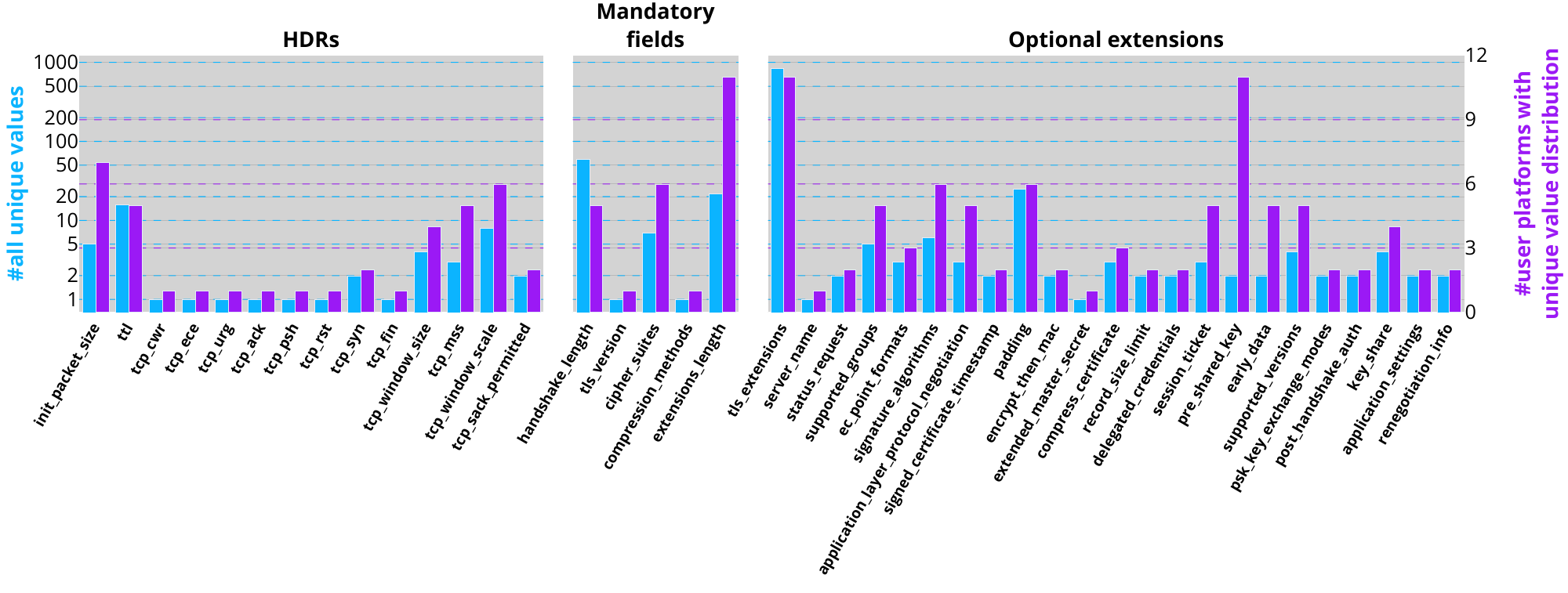}
		\label{fig:field_distribution_netflix}}
	\subfigure[Handshake field value distribution of \textbf{\color{blizzardblue}Disney+} flows.]{
		\includegraphics[width=0.47\textwidth]{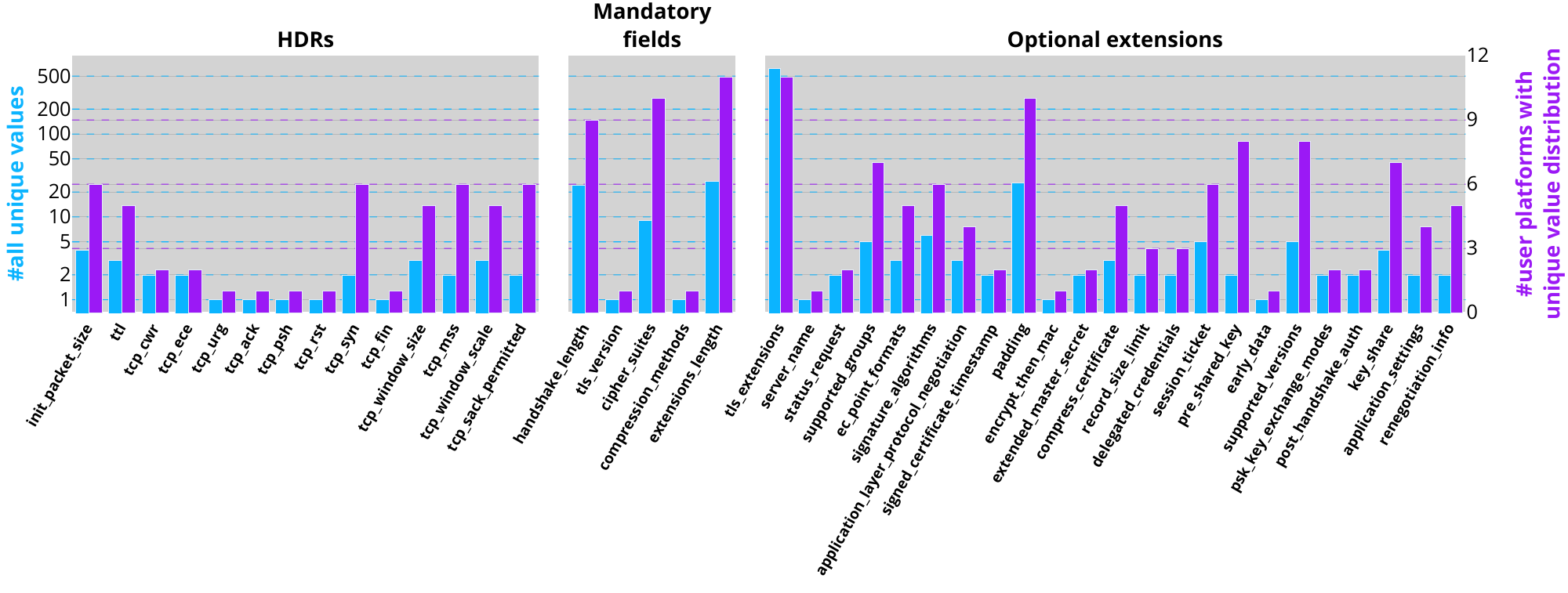}
		\label{fig:field_distribution_disney}}
	\subfigure[Handshake field value distribution of \textbf{\color{blue}Amazon Prime Video} flows.]{
		\includegraphics[width=0.47\textwidth]{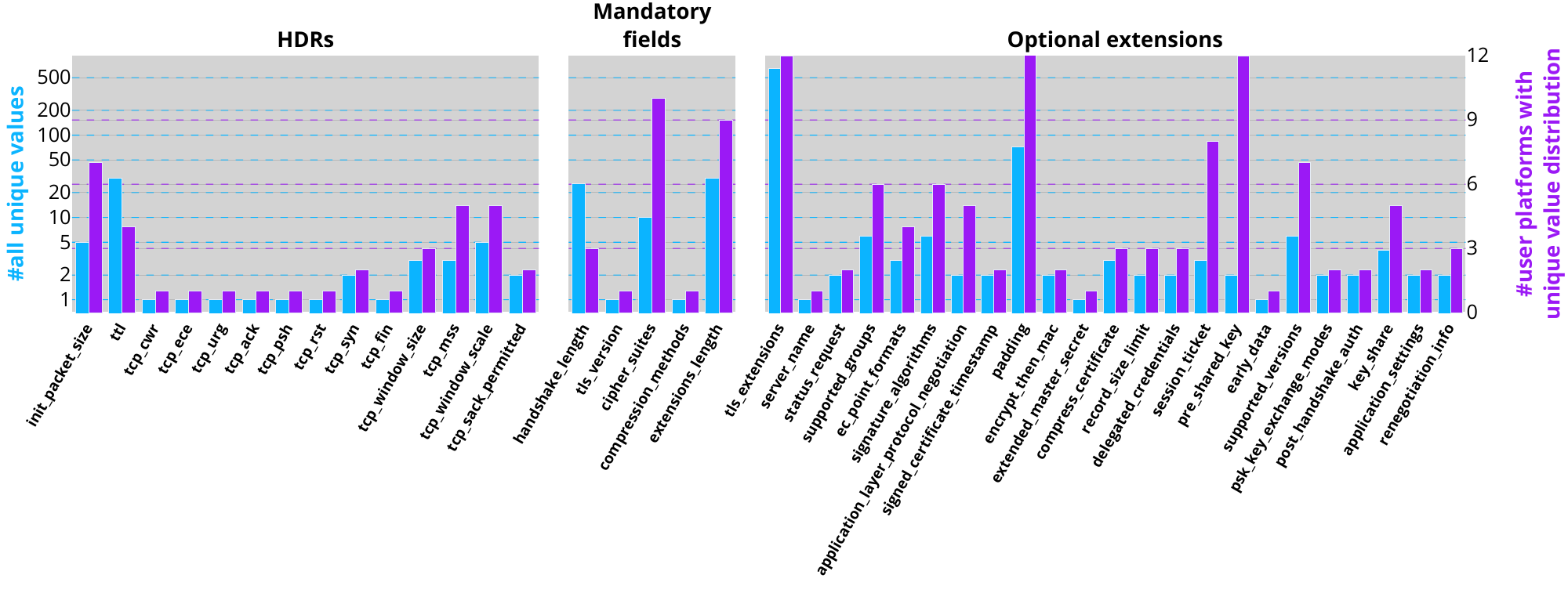}
		\label{fig:field_distribution_amazon}}
	\caption{Number of unique values (left blue) and number of user platforms with different value distributions (right purple) for each handshake field in {\color{blue-violet}Netflix}, {\color{blizzardblue}Disney+} and {\color{blue}Amazon Prime Video} flows over TCP.}
	\label{fig:field_distribution_appendix}
\end{figure}

For the other three platforms (\ie Netflix, Disney and Amazon) which delivers video flows over TCP only, we show the overall number of unique field values and number of user platforms with different value distributions for each handshake field in Fig.~\ref{fig:field_distribution_appendix}, using a similar layout as Fig.~\ref{fig:field_distribution_quic}.
For most handshake fields, the value distributions either vary significantly across user platforms (e.g., \textit{cipher\_suites}) or remain consistently stable across user platforms (e.g., \textit{compression\_methods}), regardless of the video provider. There are also exceptions such as \textit{tcp\_syn} where six unique value distributions are observed across Disney+ user platforms, whereas the value distribution of this field remains consistent across user platforms for Netflix and Amazon. Such instances suggest that the indicative power of a certain handshake fields may vary for different video providers.

\clearpage

\section{Attribute Importance}
\begin{figure}[H]
	\subfigure[Attribute importance for classifying \textbf{\color{blue-violet}Netflix} user platforms.]{
		\includegraphics[width=\columnwidth]{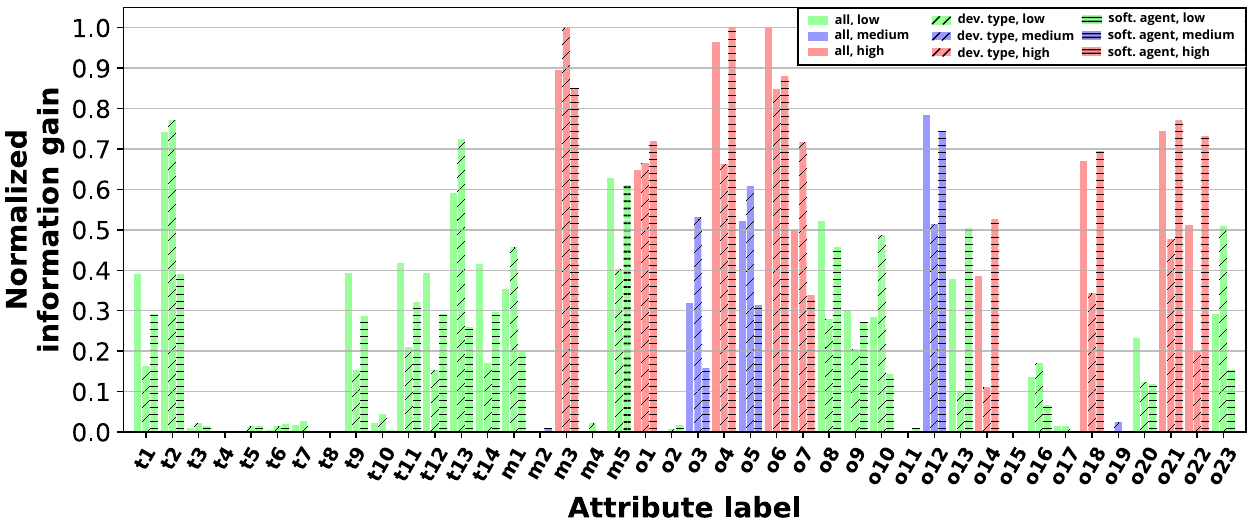}
		\label{fig:feature_importance_netflix}
	}
	\subfigure[Attribute importance for classifying \textbf{\color{blizzardblue}Disney+} user platforms.]{
		\includegraphics[width=\columnwidth]{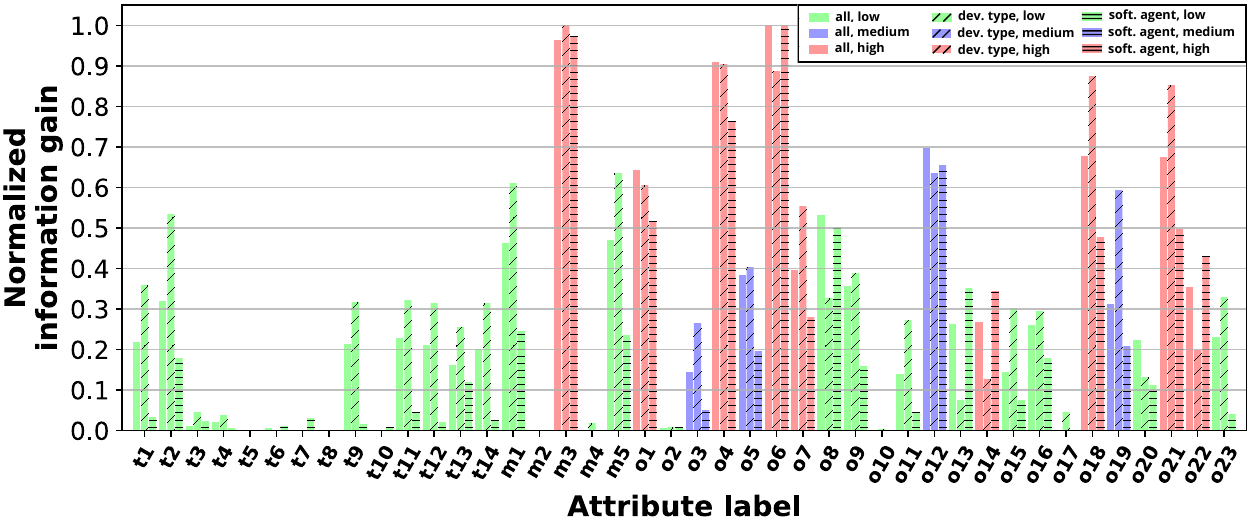}
		\label{fig:feature_importance_disney}
	}
	\subfigure[Attribute importance for classifying \textbf{\color{blue}Amazon Prime Video} user platforms.]{
		\includegraphics[width=\columnwidth]{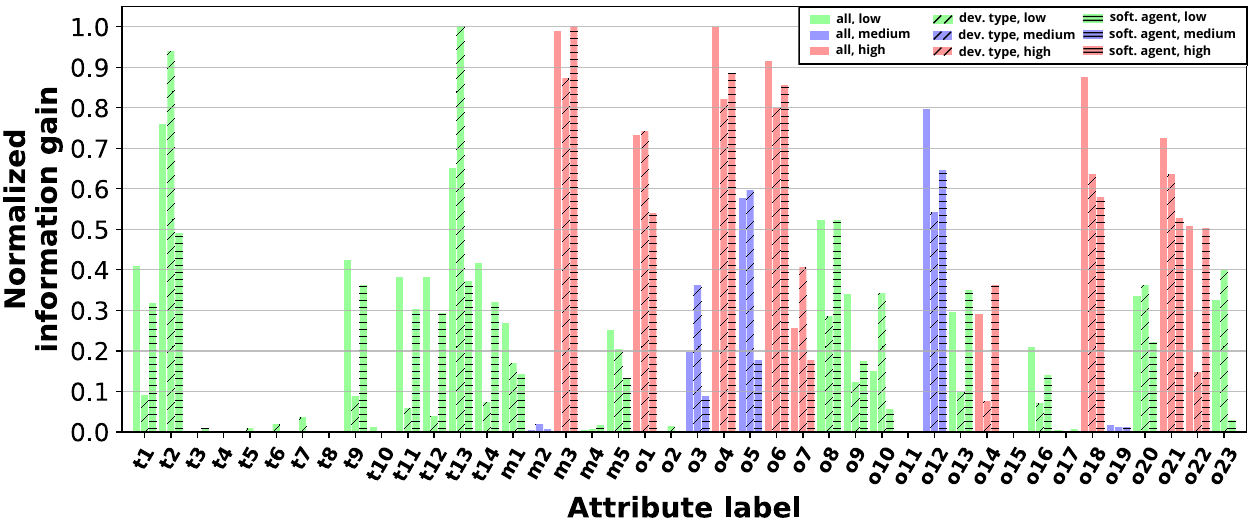}
		\label{fig:feature_importance_amazon}
	}
	\caption{Importance of different attributes in classifying user platforms of (a) {\color{blue-violet}Netflix}, (b) \textbf{\color{blizzardblue}Disney} and (c) {\color{blue}Amazon} TCP video flows.}
	\label{fig:feature_importance_others}
\end{figure}

In Fig.~\ref{fig:feature_importance_others}, we show the importance of attributes (defined in Table~\ref{tab:AppendixFeatureList}) in predicting user platforms for Netflix, Disney and Amazon Prime video flows over TCP. The attributes are color- and pattern-coded by their computational costs and prediction objectives.

We highlight that the importance (\ie normalized information gain) of a certain attribute can differ across video providers. For example, $o_{19}$ has very low importance (around 0) for all classification objectives related to Netflix and Amazon. However, it is highly useful for classifying user platforms for Disney, especially device types, where it has an importance score of nearly 0.6.
Additionally, even for attributes that are indeed of high importance across all video providers, the specific classification objectives they are useful for can differ. For instance, $t_{9}$ is of high importance for classifying device types for Disney, whereas its importance for Netflix and Amazon is medium and low, respectively. Instead, it proves to be highly useful in classifying the software agent for those other two video providers. 

\end{document}